\documentclass[aps,prb,notitlepage,superscriptaddress,floatfix,%
longbibliography,twocolumn]{revtex4-2}

\usepackage{amsmath}
\usepackage{amssymb}
\usepackage{maybemath}
\usepackage[dvipsnames]{xcolor}
\usepackage{graphicx}
\usepackage[caption=false]{subfig}
\usepackage{bm}
\usepackage{bbm}

\usepackage{siunitx}
\usepackage{soul} 

\def\hm{\hphantom{-}}

\usepackage{dcolumn}
\newcolumntype{d}[1]{D{.}{.}{#1}}
\newcolumntype{.}{D{x}{}{9}}
\newcolumntype{,}{D{x}{}{5}}
\newcolumntype{;}{D{x}{}{19}}

\usepackage{booktabs}   
\usepackage{natbib}     
\newcommand{\ii}{\mathrm{i}}
\def\dd{{\mathrm d}}
\renewcommand{\Re}{{\mathrm{Re}}}
\renewcommand{\Im}{{\mathrm{Im}}}

\newcommand{\C}{^\circ\mathrm{C}}
\newcommand{\K}{\rm K}

\usepackage{derivative}    
\derivset{\pdv}[delims-eval=.|]     

\newcommand{\CaFT}{CaF\textsubscript{2}}
\newcommand{\CaF}[1]{CaF\textsubscript{#1}}

\newcommand{\figwidthII}{0.45\textwidth} 

\def\rmC{{\mathrm{C}}}

\def\LO{{\mathrm{LO}}}
\def\TO{{\mathrm{TO}}}

\def\dd{{\mathrm d}}
\def\ii{{\mathrm i}}
\def\ee{{\mathrm e}}

\def\neff{{n_{\mathrm{eff}}}}
\def\brek{\mathrm{br}}
\def\crit{\mathrm{cr}}

\definecolor{burntorange}{rgb}{0.8, 0.33, 0.0}
\definecolor{garrosgreen}{rgb}{0.1, 0.4, 0.1}
\definecolor{dartmouthgreen}{rgb}{0.05, 0.5, 0.06}
\definecolor{duelferred}{rgb}{0.7, 0.2, 0.1}
\definecolor{cambridgeblue}{rgb}{0.1, 0.3, 1.0}
\definecolor{oxfordblue}{rgb}{0.05, 0.2, 0.7}

\newcommand{\RRCO}{{\mathrm{RRCO}}}

\newcommand{\angstrom}{\mbox{\AA}}

\begin{document}

\title{Temperature--Dependent Dielectric Function of Calcium Fluoride:\\
From a Compact Functional Form to Atom--Surface Interactions}

\author{T. Das}
\affiliation{Department of Physics and LAMOR, Missouri University of Science and
Technology, Rolla, Missouri 65409, USA}

\author{D. Alam}
\affiliation{Department of Physics and Astronomy,
University of Missouri, Columbia, Missouri 65211, USA}
\affiliation{Department of Mathematics, Physics and Computer Science, 
Lincoln Memorial University, Harrogate, TN 37752, USA}

\author{C. A. Ullrich}
\affiliation{Department of Physics and Astronomy,
University of Missouri, Columbia, Missouri 65211, USA}

\author{U. D. Jentschura}
\affiliation{Department of Physics and LAMOR, Missouri University of Science and
Technology, Rolla, Missouri 65409, USA}

\begin{abstract}
The optical properties of calcium fluoride (fluorspar, \CaFT{})
are mainly determined by a
strongly temperature-dependent giant infrared (IR) peak,
and a series of nearly temperature-independent ultraviolet (UV) peaks.
We find that the temperature dependence of the
IR peak can be modeled, to good accuracy,
by a radiation-reaction improved coupled-oscillator model (RRCO model),
with temperature-dependent parameters.
For the UV peaks, we find a convenient functional
form which covers both the real as well as the
imaginary parts of the dielectric function and provide a comparison
to first-principles calculations
based on time-dependent density-functional theory (TDDFT).
The result is a compact functional form
for the dielectric function of undoped \CaFT{}
applicable to wide frequency and temperature ranges
($0 < \hbar \omega < 60 \, \mathrm{eV}$, $22\C < T < 500\C$).
With the help of the temperature-dependent
dielectric function, we obtain temperature-dependent values of the
short-range and long-range asymptotics of atom-surface
interactions with \CaFT{},
for hydrogen, as well as ground-state and metastable helium.
The giant IR absorption peak of \CaFT{} is shown
to lead to a delayed onset of the fully retarded Casimir--Polder
limit in the long-range interaction regime.
We present arguments supporting 
a more general applicability of the RRCO model to 
materials of general interest.
\end{abstract}

\maketitle


\section{Introduction}

Due to its wide transmission band associated with a nearly transparent region
in the visible range of the optical spectrum, calcium fluoride (\CaFT{},
commonly referred to as fluorspar) has many technological applications.~{\em
E.g.}, it is used for anti-reflective coatings on lenses~\cite{Ba1946},
transparent substrates in photolithography~\cite{BlEtAl1997}, as a
substrate material for infrared spectroscopy~\cite{MaPaHuPo2020}, and
as a medium for supercontinuum generation~\cite{JoPaMi2009}. 
The material
is subjected to extreme thermal conditions in instances such as optical
elements in high-energy lasers~\cite{Kl2006}. Other environments with extreme
conditions include the Near--Infrared Spectrometer and Photometer (NISP) on the
Euclid satellite~\cite{LeEtAl2015temp}.
A very illustrative discussion of the utility
of \CaFT{} for 193nm-lithography (together
with an illustrative discussion of the
favorable properties of the material against laser-induced damage)
is given in Ref.~\cite{Ma2008CaF2}.
Accurate knowledge of the optical
and dielectric properties of \CaFT{} and their temperature dependence is thus
beneficial for these applications. A convenient functional form, to describe
the dielectric function over wide ranges of temperature and frequency, is also
required and desirable for the calculation
of atom-surface interactions and quantum-reflection
studies.

The structure of the dielectric function of \CaFT{} is characterized by
a giant infrared (IR) resonance, where the
maximum modulus of the
real part of the dielectric function
exceeds the value of $|\Re(\epsilon)| > 80$,
while rapidly changing its sign at the resonance,
and a series of peaks in the ultraviolet (UV).
The IR peak has recently been
analyzed in Ref.~\cite{PSEtAl2009}, while
the transparent region within the visible range
has been analyzed in relatively recent
articles~\cite{Li1980halides,DaMa2002,FiSaEnSu2008,LeEtAl2015temp,KGEtAl2017,ZhWaTh2023}.
By contrast, data collected
for the IR region in Ref.~\cite{Be1985},
while capturing the main characteristics of the
frequency-dependent dielectric function,
must be considered outdated at the current stage. For the UV region,
we use data from Refs.~\cite{To1936,LeStRo1965}.
The data collected in Ref.~\cite{LeStRo1965}
display five overlapping resonances
in the UV range $0.35 \, {\rm a.u.} < \omega < 1.3 \, {\rm a.u.}$
(here, ``a.u.'' stands for atomic units,
where the photon energy is measured in in units of the Hartree energy,
see Chap.~2 of Ref.~\cite{JeAd2022book}).
It is essential that the data from Ref.~\cite{LeStRo1965}
display clear excitonic signatures at a
UV photon energy of $\hbar \omega \approx 10.8$\,eV.
Namely, apart from the fact that excitonic effects dominate the optical properties
of wide-gap insulators close to the optical gap,
these features are also essential for the comparison of experimental
data to first-principles calculations, which
are reported here using
time-dependent density-functional theory (TDDFT).

We have recently investigated the dielectric functions
of intrinsic silicon~\cite{MoEtAl2022,MoEtAl2025erratum}
and $\alpha$-quartz~\cite{Je2024multipole}
and found that a radiation-reaction improved coupled-oscillator model
(RRCO model) is very well suited
for the description of both real and
imaginary parts of the dielectric function.
The justification for the RRCO model
was discussed in Ref.~\cite{DaUlJe2025coupled},
on the basis of model calculations with damped, coupled
oscillators, which represent a natural extension
of the sum over uncoupled oscillators which
form the basis of the original Sellmeier model~\cite{Se1872}.
The recent investigation~\cite{DaUlJe2025coupled}
thus provides additional justification
for the very similar functional forms (which all
involve complex rather than real oscillator strengths) that have been
discussed in Eq.~(1) of Ref.~\cite{Ri1985} for rutile, in
Eq.~(4) of Ref.~\cite{Tr1985} for cubic thallium, and in
Eq.~(1) of Ref.~\cite{PaKh1985} for sodium nitrate.
Also, we mention the recent applications
to intrinsic silicon and $\alpha$-quartz discussed
in Refs.~\cite{MoEtAl2022,Je2024multipole,MoEtAl2025erratum}.
Hence, it is indicated to investigate
if the same analytic model,
which has a comparatively simple analytic form,
can be successfully applied
to other technologically important substances
such as \CaFT{}. As a byproduct of our investigations,
we shall thus aim to update and improve
the four-term Sellmeier model that is
otherwise used in Ref.~\cite{DaMa2002}
to describe the experimental data for \CaFT{}.

Here, we confirm that, although a
traditional Sellmeier-type model is well suited
in representing the optical data
in the transparent region, it fails to represent the complicated
structure found in the UV region, and in the IR region.
Our first aim is to find a unified functional
form which, for room temperature,
covers the entire spectral range from the IR to the UV.
Our second aim is to describe the temperature dependence
of the dielectric function.
In Ref.~\cite{PSEtAl2009}, the temperature dependence
of the IR peak was described using an elaborate functional form.
The temperature dependence in
the visible region has been analyzed
by temperature-dependent Sellmeier functional forms
in Refs.~\cite{LeEtAl2015temp} and~\cite{ZhWaTh2023}.
Our aim is to represent the temperature dependence of the dielectric
function using an economic functional form that captures the temperature
dependence in the IR and visible region and also reproduces the peaks in the UV
region which are mostly temperature independent,
and to confront the results with first-principles calculations
in the UV region.

This paper is organized as follows.
In Sec.~\ref{sec2A} we analyze, very briefly, the
structure of the IR and UV resonances of \CaFT{}.
The IR peak, including its temperature dependence,
is analyzed in Sec.~\ref{sec2B}.
The UV peaks are analyzed in Sec.~\ref{sec2C}.
Results for the temperature- and frequency-dependent
dielectric function are discussed in Sec.~\ref{sec2D}.
In Sec.~\ref{sec3}
we provide additional insight into the UV response of CaF$_2$ through TDDFT.
Atom-surface potentials are calculated in Sec.~\ref{sec4}.
Conclusions are reserved for Sec.~\ref{sec5}.
A statistical analysis of the fit is described in
Appendix~\ref{appa}, and
an illustrative discussion of the natural cleavage
planes, of the crystal symmetry groups,
and of other limiting factors,
is contained in Appendix~\ref{appb}.
Appendix~\ref{appc} is devoted to a comparison with
available data sheets~\cite{CaF2hellma,CaF2corning}.
Finally, in Appendix~\ref{appd}, we 
investigate the dielectric function of rutile,
based on the RRCO model.

\begin{table}[t!]
\centering
\caption{ \label{table1}
Parameters for fitting the infrared (IR)
peak at different temperatures while keeping the ultraviolet (UV) parameters
fixed. For finding the temperature-dependent fits, we have used data from
Ref.~\cite{PSEtAl2009} for the IR peak,
Ref.~\cite{ZhWaTh2023} for the transparent region, and UV data
from Ref.~\cite{LeStRo1965}. Results are given in atomic units ($E_h$ is the 
Hartree energy).}
\setlength{\tabcolsep}{3.5pt}
\begin{tabular}{lllll}
\toprule
\toprule
\multicolumn{1}{c}{$T$ [$\C$]} &
\multicolumn{1}{c}{$a_1$} & \multicolumn{1}{c}{$\omega_1~[E_h/\hbar]$} &
\multicolumn{1}{c}{$\gamma_1~[E_h/\hbar]$} & 
\multicolumn{1}{c}{$\gamma^\prime_1~[E_h/\hbar]$} \\
\midrule
22  & $4.311$ & $1.184 \times 10^{-3}$ & $3.025 \times 10^{-5}$ & $3.990 \times 10^{-6}$ \\
100 & $4.365$ & $1.169 \times 10^{-3}$ & $4.220 \times 10^{-5}$ & $1.199 \times 10^{-5}$ \\
200 & $4.520$ & $1.148 \times 10^{-3}$ & $6.343 \times 10^{-5}$ & $6.140 \times 10^{-5}$ \\
300 & $4.815$ & $1.124 \times 10^{-3}$ & $1.042 \times 10^{-4}$ & $8.496 \times 10^{-5}$ \\
400 & $5.105$ & $1.100 \times 10^{-3}$ & $1.703 \times 10^{-4}$ & $6.712 \times 10^{-5}$ \\
500 & $5.687$ & $1.088 \times 10^{-3}$ & $2.837 \times 10^{-4}$ & $5.422 \times 10^{-5}$ \\
\bottomrule
\bottomrule
\end{tabular}
\end{table}

\begin{table}[t!]
\centering
\caption{ \label{table2}
Temperature coefficients found by fitting
IR parameters in Table~\ref{table1} using Eq.~\eqref{IRtempFit}.
Results for $C_j^{(\omega_1)}$,
$C_j^{(\gamma_1)}$, and
$C_j^{(\gamma'_1)}$ are given in units
of $E_h/\hbar$, while the 
$C_j^{(a_1)}$ are dimensionless ($j=0,1,2,3$).
}
\setlength{\tabcolsep}{0.8pt}
\squeezetable
\begin{tabular}{lllll}
\toprule
\toprule
& 
\multicolumn{1}{c}{$C_0$} & 
\multicolumn{1}{c}{$C_1$} & 
\multicolumn{1}{c}{$C_2$} &
\multicolumn{1}{c}{$C_3$} \\
\midrule
$a_1$             & $4.301 \times 10^0$       & $\hm  2.401 \times 10^{-1}$  & $\hm 1.215 \times 10^{-1}$ & $\hm 1.492 \times 10^{-1}$ \\
$\omega_1$        & $1.183 \times 10^{-3}$    &     $-3.783 \times 10^{-5}$  & $   -4.641 \times 10^{-5}$ & $\hm 2.062 \times 10^{-5}$ \\
$\gamma_1$        & $2.966 \times 10^{-5}$    & $\hm  4.973 \times 10^{-5}$  & $   -2.297 \times 10^{-5}$ & $\hm 5.320 \times 10^{-5}$ \\
$\gamma^\prime_1$ & $3.990 \times 10^{-6}$    & $\hm  6.414 \times 10^{-5}$  & $\hm 5.962 \times 10^{-5}$ &    $-4.974 \times 10^{-5}$ \\
\bottomrule
\bottomrule
\end{tabular}
\end{table}

\begin{table}[t!]
\centering
\caption{\label{table3}
Parameters for fitting the UV range of the dielectric
function, $\epsilon(\omega)$ against the data from
Refs.~\cite{KGEtAl2017,Li1980halides,LeStRo1965}.}
\setlength{\tabcolsep}{2.5pt}
\begin{tabular}{cllll}
\toprule
\toprule
\multicolumn{1}{c}{$k$} & 
\multicolumn{1}{c}{$a_k$} & 
\multicolumn{1}{c}{$\omega_k~[E_h/\hbar]$} &
\multicolumn{1}{c}{$\gamma_k~[E_h/\hbar]$} & 
\multicolumn{1}{c}{$\gamma^\prime_k~[E_h/\hbar]$} \\
\midrule
2 & $6.332\times 10^{-2}$ & $4.040\times 10^{-1}$ & $1.918\times 10^{-2}$ &
  $\hphantom{-}3.259\times 10^{-1}$ \\
3 & $1.264\times 10^{-1}$ & $4.737\times 10^{-1}$ & $2.835\times 10^{-2}$ &
  $-9.653\times 10^{-2}$ \\
4 & $4.704\times 10^{-1}$ & $5.321\times 10^{-1}$ & $1.003\times 10^{-1}$ &
  $\hphantom{-} 2.151\times  10^{-1}$ \\
5 & $3.393\times 10^{-1}$ & $1.043\times 10^0$    & $4.762\times 10^{-1}$ &
  $-4.610\times  10^{-1}$ \\
6 & $4.173\times 10^{-2}$ & $1.234\times 10^0$    & $9.917\times 10^{-2}$ &
  $\hphantom{-} 1.813\times 10^0$  \\
\bottomrule
\bottomrule
\end{tabular}
\end{table}

\begin{figure}[t!]
\begin{center}
\includegraphics[width=0.91\linewidth]{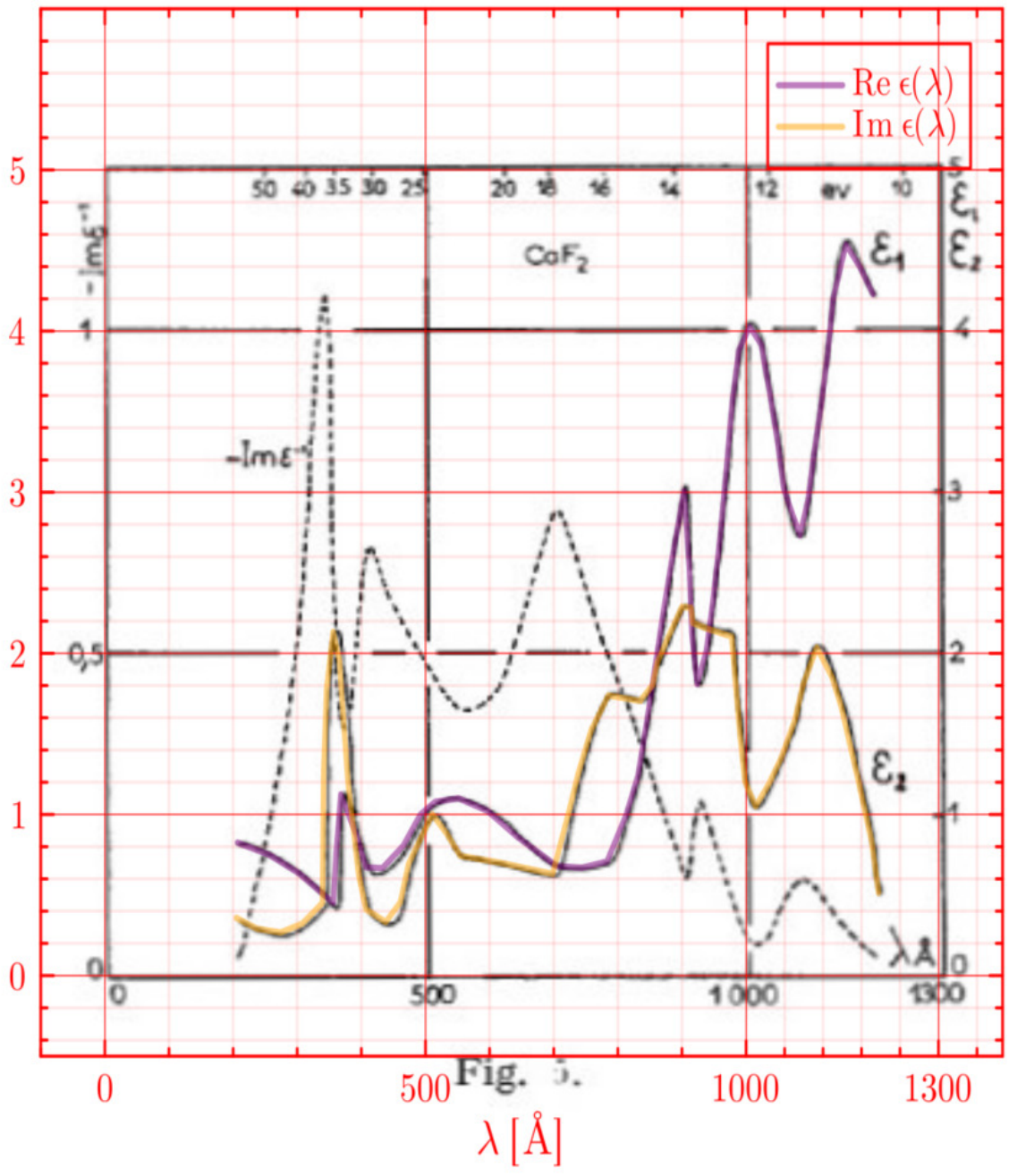}
\caption{ \label{fig1}
Experimental data of the dielectric function of \CaFT{}{} single crystals
versus wavelength $\lambda$
(adapted with permission~\cite{cr_clearance}
from Fig.~5 of Ref.~\cite{LeStRo1965}).
Five resonances are clearly identified. The authors of Ref.~\cite{LeStRo1965}
use the identification $\varepsilon_1 \equiv \Re(\epsilon)$
and $\varepsilon_2 \equiv \Im(\epsilon)$.
The extracted purple curve (real part) and the yellow
curve (imaginary part) are
superimposed on the numerical data of Ref.~\cite{LeStRo1965}.}
\end{center}
\end{figure}

\begin{figure*}[t!]
\begin{center}
\begin{minipage}{0.7\linewidth}
\begin{center}
\includegraphics[width=0.91\linewidth]{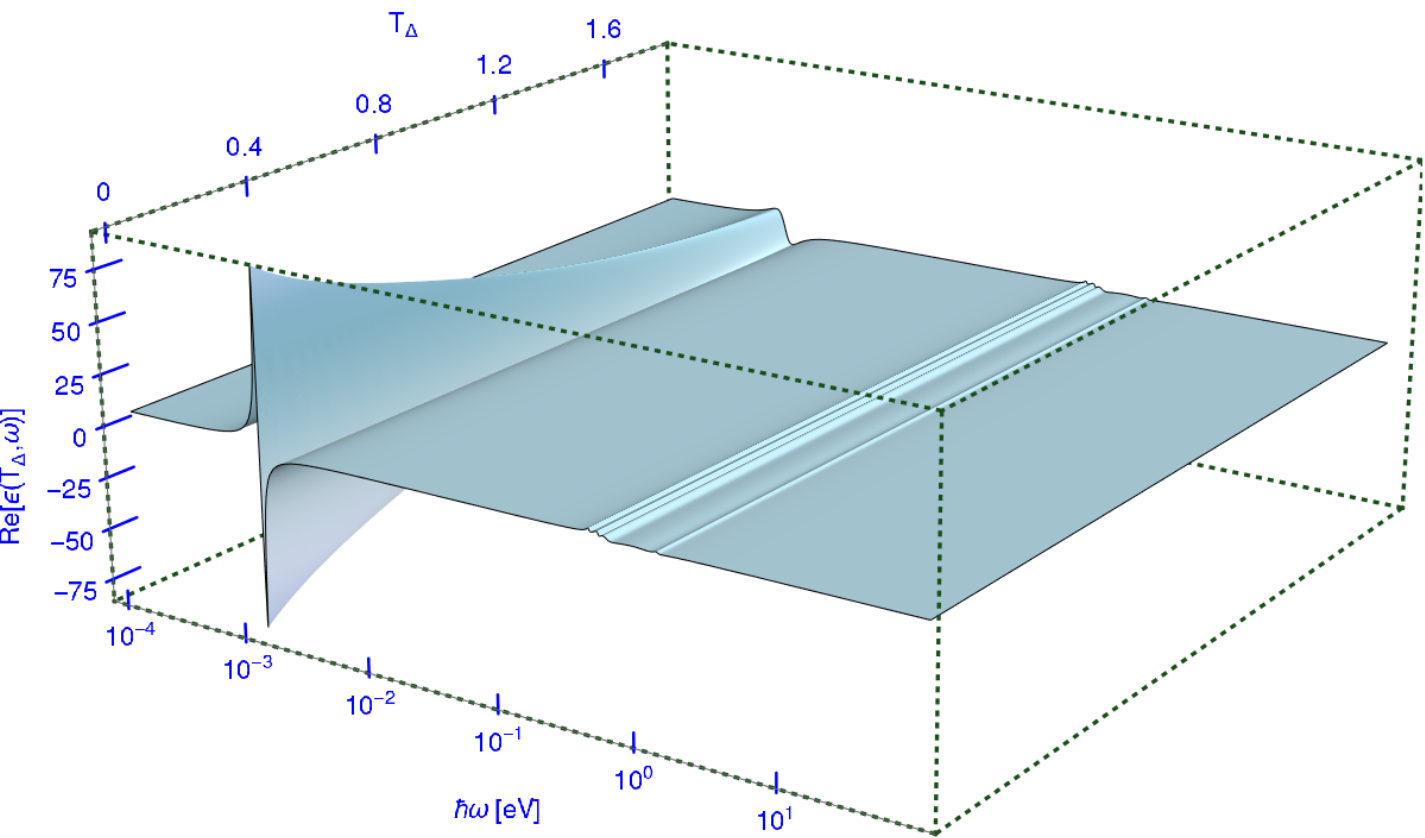}
\caption{\label{fig2}
Real part of the dielectric
function ${\rm Re}[\epsilon(\omega)]$,
plotted using the temperature coefficients from
Table~\ref{table2} for the IR resonance and the UV parameters from
Table~\ref{table3}.  }
\end{center}
\end{minipage}
\end{center}
\end{figure*}

\begin{figure*}[t!]
\begin{center}
\begin{minipage}{0.7\linewidth}
\begin{center}
\includegraphics[width=0.91\linewidth]{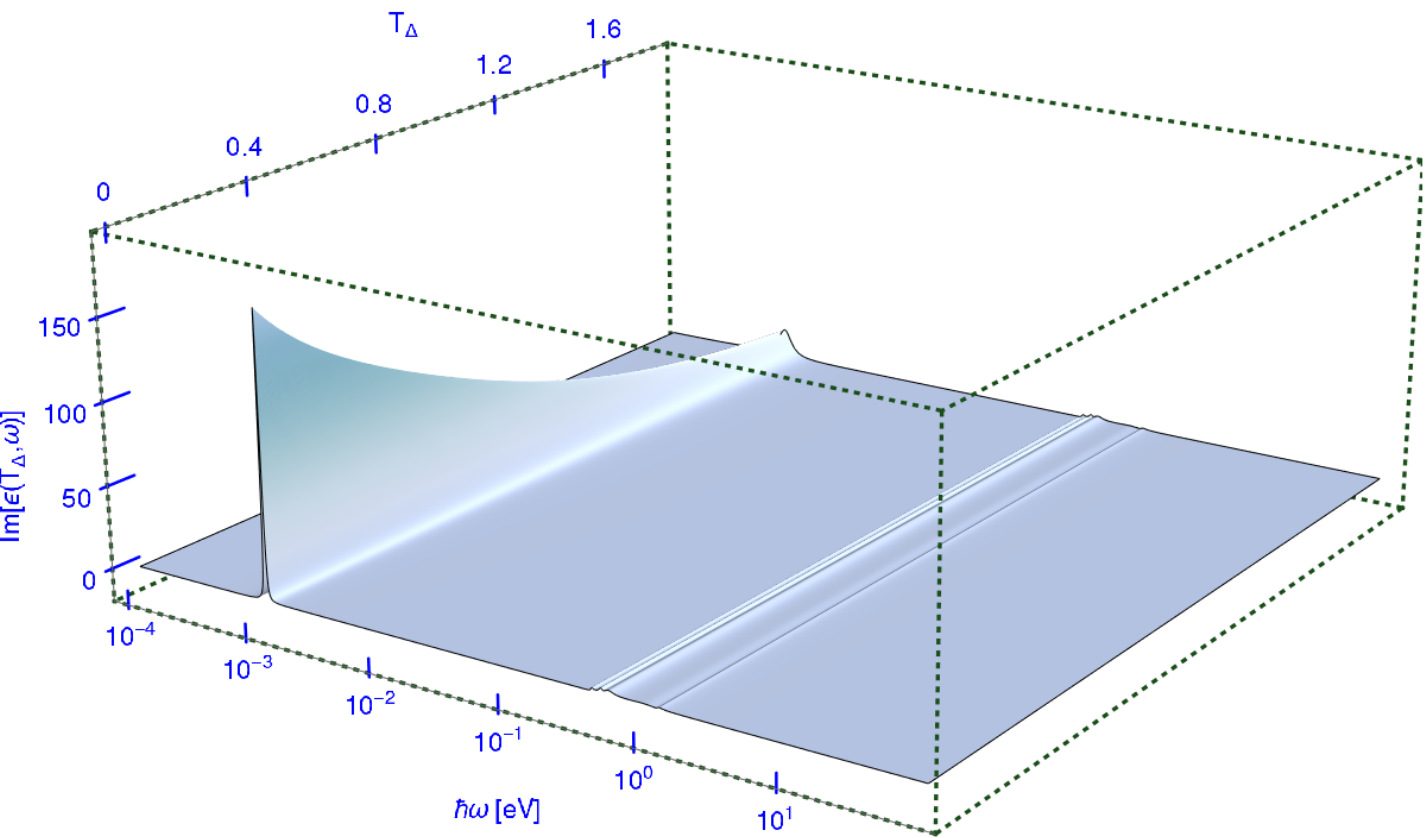}
\caption{\label{fig3}
Same as Fig.~\ref{fig2},
but for the imaginary part of the dielectric function.}
\end{center}
\end{minipage}
\end{center}
\end{figure*}

%
%
\section{Dielectric Function of C\lowercase{a}F{\textsubscript{2}}}
\label{sec2}

%
%
\subsection{Resonances}
\label{sec2A}

A careful examination of the experimental data obtained
in Refs.~\cite{PSEtAl2009,Li1980halides,DaMa2002,%
FiSaEnSu2008,LeEtAl2015temp,KGEtAl2017,ZhWaTh2023,%
To1936,LeStRo1965} reveals the presence
of one dominant IR resonance and a total of five UV resonances
(see Fig.~\ref{fig1}).
The RRCO model is thus used in the form
\begin{equation}
\label{master}
\epsilon_{\RRCO{}}(T_\Delta, \omega) = 1 + \sum_{k=1}^{k_{\rm max} = 6}
\frac{a_k(\omega_k^2 - \ii \gamma_k'\omega)}%
{\omega_k^2 - \omega^2 - \ii \omega \gamma_k} \,,
\end{equation}
with temperature-dependent parameters $a_k$,
$\omega_k$, $\gamma_k$ and $\gamma'_k$. Here,
\begin{equation}
\label{defTdelta}
T_\Delta = \frac{T-T_0}{T_0}, \qquad T_0 = 293\K \,,
\end{equation}
is the reduced temperature.
The value of $T_0 = 293\,{\rm K}$
(as opposed to, say, $T_0 = 295\,{\rm K}$) is
used here for room temperature, in accordance with
Refs.~\cite{MoEtAl2022,MoEtAl2025erratum}.

The identification is as follows:
The $\omega_k$ are the resonance frequencies,
the $\gamma_k$ are the width parameters,
the $\gamma'_k$ are the parameters of our model,
and the parameters $a_k$ represent the amplitudes.
The relation of the RRCO model
given in Eq.~\eqref{master} to
coupled oscillators and radiative reaction
is discussed in Ref.~\cite{DaUlJe2025coupled}.

\begin{figure*}[t!]
\begin{minipage}{0.93\linewidth}
\begin{center}
\includegraphics[width=0.93\linewidth]{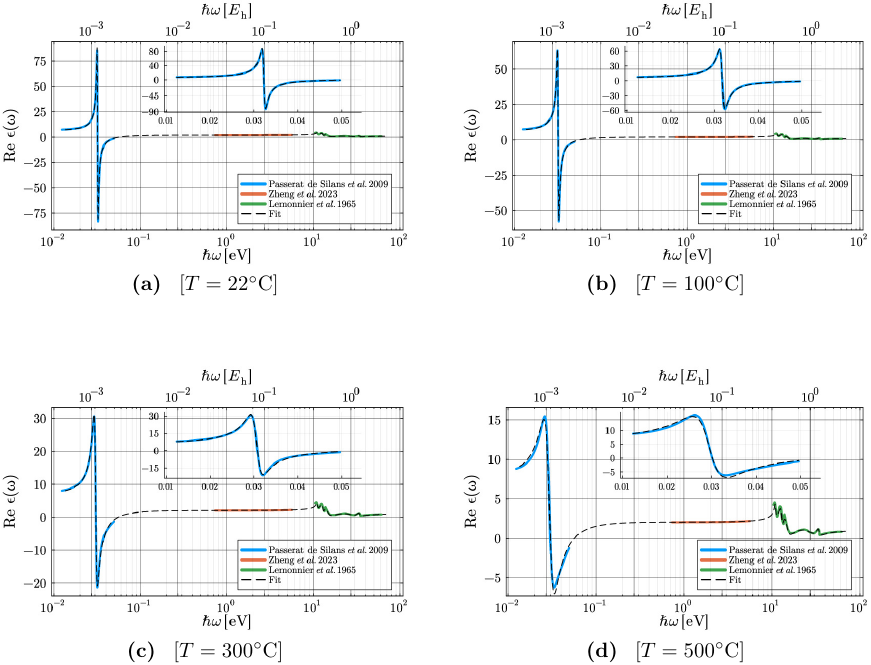}
\caption{\label{fig4}
The real part of the dielectric function,
${\rm Re}[\epsilon(\omega)]$ is shown at different temperatures:
panel (a) for $T = 22\, {}^\circ\rmC$,
panel (b) for $T = 100 \, {}^\circ\rmC$,
panel (c) for $T = 300 \, {}^\circ\rmC$, and
panel (d) for $T = 500 \, {}^\circ\rmC$.
The IR parameters from Table~\ref{table1} and
the UV parameters from Table~\ref{table3} are employed. The black dashed line
represents the fit; the blue, red and green curves are from
Ref.~\cite{PSEtAl2009} (Passerat de Silans {\em et al.}),
Ref.~\cite{ZhWaTh2023} (Zheng {\em et al.}),
and Ref.~\cite{LeStRo1965} (Lemonnier {\em et al.}), respectively.}
\end{center}
\end{minipage}
\end{figure*}

%
%
\subsection{Giant IR Peak}
\label{sec2B}

The very careful investigation in Ref.~\cite{PSEtAl2009}
is our primary resource for temperature-dependent
data on the frequency-dependent
dielectric function around the
IR resonance at $\omega \approx 1.18 \times 10^{-3} \, {\rm a.u.}$.
However, even if one were to restrict the discussion
to the IR region alone, it would be inconsistent to
obtain the fitting parameters for the
coupled-oscillator model given in Eq.~\eqref{master} based on IR
data alone, because the functional form
given in Eq.~\eqref{master} is not restricted
to a particular range of frequency values,
and a change in the fitting parameters for
the IR region may thus influence the behavior of $\epsilon(\omega)$ in
the UV region as well.
Thus, in order to allow for a smooth transition to the transparent (visible)
and UV region, we use temperature-dependent data from
Refs.~\cite{PSEtAl2009,ZhWaTh2023} in order to fit the IR parameters, {\em i.e.},
the temperature-dependent parameters $a_1$, $\omega_1$, $\gamma_1$ and
$\gamma^\prime_1$, in the temperature range $22\C < T < 500\C$.
For room temperature, we have checked the consistency
of the used data sets with Refs.~\cite{KGEtAl2017,Li1980halides,DaMa2002,FiSaEnSu2008}.
Results are communicated in Table~\ref{table1}.
For the parameters, we use atomic units
(see Chap.~2 of Ref.~\cite{JeAd2022book}),
{\em i.e.}, the unit of energy is the Hartree energy,
$E_h = 27.2114 \, {\rm eV}$, the unit of length is the Bohr radius,
$a_0 = 0.529177 \, \angstrom$. We recall that,
in the atomic unit system,
$\hbar = 1$, $\epsilon_0 = 1/(4 \pi)$, and
$c = 1/\alpha$ (where $\alpha$ is the fine-structure constant).
The unit of charge is the elementary charge
(equal to the modulus of the electron charge).
The atomic unit of frequency is $E_h/h = 6.57968 \times 10^{15} \, {\rm Hz}$,
and the atomic unit of angular frequency is $E_h/\hbar 
= 2 \pi \times 6.57968 \times 10^{15} \, {\rm Hz}
= 4.13414 \times 10^{16} \, {\rm rad}/{\rm s}$.

In terms of the reduced temperature $T_\Delta$
defined in Eq.~\eqref{defTdelta},
we find that the IR parameters for different temperatures
can be expressed succinctly with a cubic polynomial,
as follows,
\begin{subequations}
\label{IRtempFit}
\begin{align}
a_1 =& \; C_0^{(a_1)} + C_1^{(a_1)} \, T_\Delta +
C_2^{(a_1)} T_\Delta^2 + C_3^{(a_1)} \, T_\Delta^3 \,,
\\
\omega_1 =& \; C_0^{(\omega_1)} + C_1^{(\omega_1)} \, T_\Delta +
C_2^{(\omega_1)} T_\Delta^2 + C_3^{(\omega_1)} \, T_\Delta^3 \,,
\\
\gamma_1 =& \; C_0^{(\gamma_1)} + C_1^{(\gamma_1)} \, T_\Delta +
C_2^{(\gamma_1)} T_\Delta^2 + C_3^{(\gamma_1)} \, T_\Delta^3 \,,
\\
\gamma'_1 =& \; C_0^{(\gamma'_1)} + C_1^{(\gamma'_1)} \, T_\Delta +
C_2^{(\gamma'_1)} T_\Delta^2 + C_3^{(\gamma'_1)} \, T_\Delta^3 \,.
\end{align}
\end{subequations}
Numerical results are presented in Table~\ref{table2}.
In contrast to Refs.~\cite{MoEtAl2022,MoEtAl2025erratum},
we find that quadratic polynomials in $T_\Delta$
are insufficient to capture the observed temperature
dependence; third-order polynomials lead to a
satisfactory fit.

\begin{figure*}[t!]
\begin{minipage}{0.93\linewidth}
\begin{center}
\includegraphics[width=0.93\linewidth]{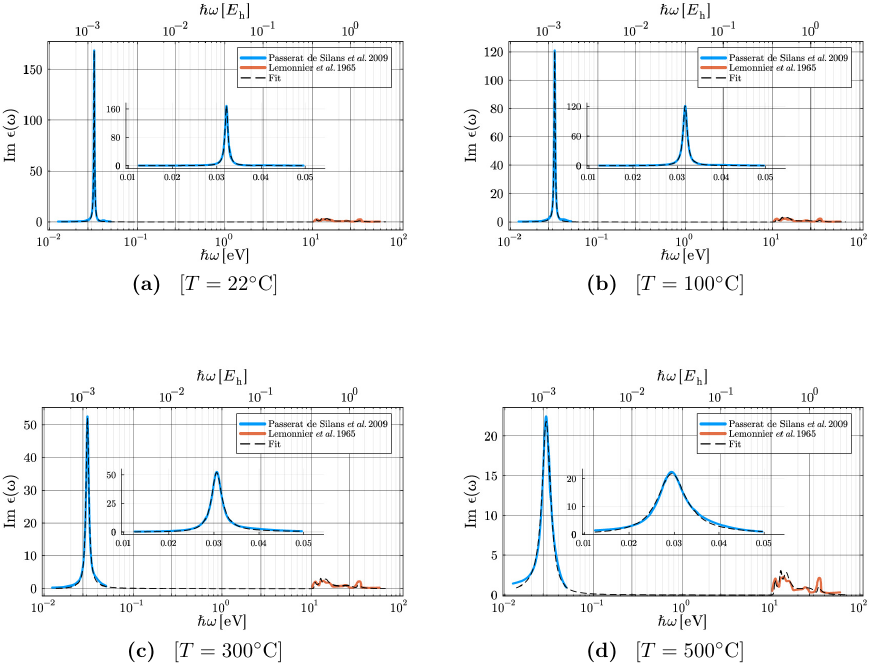}
\caption{\label{fig5}
Same as Fig.~\ref{fig4},
but for the imaginary part of the dielectric function.}
\end{center}
\end{minipage}
\end{figure*}

\begin{figure}[t!]
\begin{center}
\includegraphics[width=0.91\linewidth]{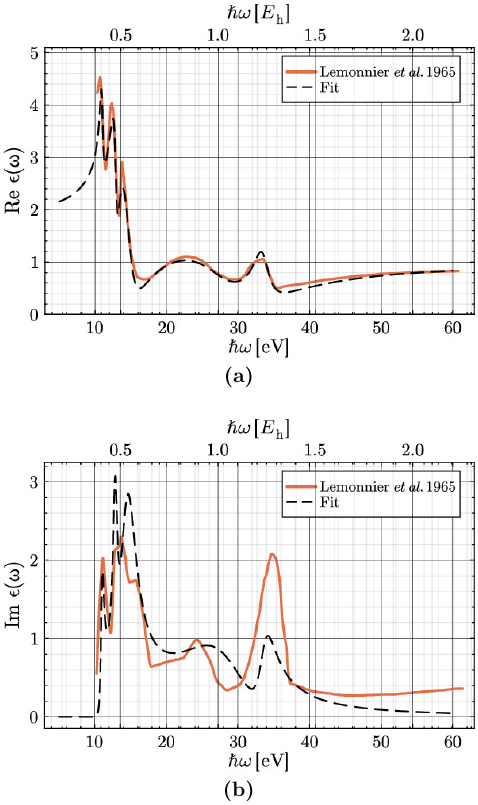}
\caption{\label{fig6}Real part (a) and
imaginary part (b) of the dielectric function $\epsilon(\omega)$
in the UV region. The dashed curve corresponds to the fit at
$T=22\C$, using data from
Tables~\ref{table1},~\ref{table2}, and~\ref{table3}.
The solid red curve represents the data compiled
in Ref.~\cite{LeStRo1965}, with data points
extracted from Fig.~5 of Ref.~\cite{LeStRo1965}.}
\end{center}
\end{figure}

%
%
\subsection{UV Peaks and Transparent Region}
\label{sec2C}

For the UV region, the main source of data is from Ref.~\cite{LeStRo1965}
(see also Fig.~\ref{fig1}).  In the visible (transparent) region,
there are no discernible resonances.
Hence, for temperature-dependent high-precision data in the transparent region,
we use the most recent, high-precision data set from Ref.~\cite{ZhWaTh2023},
which is more detailed than the data set provided
in Ref.~\cite{LeEtAl2015temp} and, as we have checked, consistent with
the data sets of Refs.~\cite{FiSaEnSu2008,KGEtAl2017,Li1980halides,DaMa2002},
which had been obtained for temperatures very close to room temperature.
The inclusion of the data sets relevant
to the transparent region ensures a smooth transition into the
visible optical range. The coefficients for $2 \leq k \leq 6$
resulting from the nonlinear fit are reported in Table~\ref{table3}.

For the UV peaks, we assume a temperature-independent model,
for reasons described in the following.
Temperature-dependent data for the dielectric function have been
obtained in Refs.~\cite{KGEtAl2017,ZhWaTh2023},
for the intermediate range
corresponding to the visible spectrum.
In this region, \CaFT{} is largely transparent.
From an inspection of the data presented in
Refs.~\cite{KGEtAl2017,ZhWaTh2023}, it becomes clear that,
near the transition to the UV,
both the real as well as the imaginary parts
of the dielectric function change by less than a percent
between 273\,K and 773\,K ($0\,\C$ and $500\,\C$).
For the transition regime to the UV,
our assumed 1\,\% uncertainty is thus larger than the
observed temperature variation in that same transition region.
We are thus safe to assume that, within the accuracy of our fit,
the temperature dependence in the UV can be ignored.
This conjecture is tied to the fact that the UV resonances arise from
electronic interband transitions, and it appears that
the band gap is virtually independent of thermal effects.
By contrast, the IR peaks are related to
vibrational excitations of the crystal lattice.

\begin{table*}[!t]
\caption{\label{table4}
Parameter values for the RRCO model
given in~Eq.~\eqref{master},
for \CaFT{}, at a temperature of $T = 22\C$.
Six resonances are identified ($k_{\rm max} = 6$).}
\centering
\setlength{\tabcolsep}{3pt}
\squeezetable
  \begin{tabular}{l l l l l}
  \toprule
  \toprule
  \multicolumn{1}{c}{$k$}
  & \multicolumn{1}{c}{$a_k$}
  & \multicolumn{1}{c}{$\omega_k~[E_h/\hbar]$}
  & \multicolumn{1}{c}{$\gamma_k~[E_h/\hbar]$}
  & \multicolumn{1}{c}{$\gamma'_k~[E_h/\hbar]$} \\
  \midrule
1 & $    4.311 \times 10^{0\hm}    \pm 3.687 \times 10^{-3}$
  & $    1.184 \times 10^{-3}    \pm 1.841 \times 10^{-8}$
  & $    3.025 \times 10^{-5}    \pm 3.615 \times 10^{-8}$
  & $\hm 3.990 \times 10^{-6}    \pm 1.050 \times 10^{-6}$ \\
  & \multicolumn{1}{r}{(0.08552  \%)}
  & \multicolumn{1}{r}{(0.001555  \%)}
  & \multicolumn{1}{r}{(0.1195  \%)}
  & \multicolumn{1}{r}{(26.31  \%)} \\[2.1ex]
2 & $    6.332 \times 10^{-2}    \pm 1.764 \times 10^{-2}$
  & $    4.040 \times 10^{-1}    \pm 1.953 \times 10^{-3}$
  & $    1.918 \times 10^{-2}    \pm 3.230 \times 10^{-3}$
  & $\hm 3.259 \times 10^{-1}    \pm 1.473 \times 10^{-1}$ \\
  & \multicolumn{1}{r}{(27.87  \%)}
  & \multicolumn{1}{r}{(0.4833  \%)}
  & \multicolumn{1}{r}{(16.85  \%)}
  & \multicolumn{1}{r}{(45.18  \%)} \\[2.1ex]
3 & $    1.264 \times 10^{-1}    \pm 2.816 \times 10^{-2}$
  & $    4.737 \times 10^{-1}    \pm 2.476 \times 10^{-3}$
  & $    2.835 \times 10^{-2}    \pm 4.370 \times 10^{-3}$
  & $    -9.653 \times 10^{-2}    \pm 1.162 \times 10^{-1}$ \\
  & \multicolumn{1}{r}{(22.28  \%)}
  & \multicolumn{1}{r}{(0.5226  \%)}
  & \multicolumn{1}{r}{(15.41  \%)}
  & \multicolumn{1}{r}{(120.3  \%)} \\[2.1ex]
4 & $    4.704 \times 10^{-1}    \pm 8.283 \times 10^{-2}$
  & $    5.321 \times 10^{-1}    \pm 6.053 \times 10^{-3}$
  & $    1.003 \times 10^{-1}    \pm 1.334 \times 10^{-2}$
  & $\hm 2.151 \times 10^{-1}    \pm 8.029 \times 10^{-2}$ \\
  & \multicolumn{1}{r}{(17.61  \%)}
  & \multicolumn{1}{r}{(1.138  \%)}
  & \multicolumn{1}{r}{(13.3  \%)}
  & \multicolumn{1}{r}{(37.33  \%)} \\[2.1ex]
5 & $    3.393 \times 10^{-1}    \pm 1.023 \times 10^{-1}$
  & $    1.043 \times 10^{0\hm}    \pm 7.590 \times 10^{-2}$
  & $    4.762 \times 10^{-1}    \pm 1.759 \times 10^{-1}$
  & $    -4.610 \times 10^{-1}    \pm 3.545 \times 10^{-1}$ \\
  & \multicolumn{1}{r}{(30.14  \%)}
  & \multicolumn{1}{r}{(7.277  \%)}
  & \multicolumn{1}{r}{(36.93  \%)}
  & \multicolumn{1}{r}{(76.9  \%)} \\[2.1ex]
6 & $    4.173 \times 10^{-2}    \pm 4.464 \times 10^{-2}$
  & $    1.234 \times 10^{0\hm}    \pm 2.472 \times 10^{-2}$
  & $    9.917 \times 10^{-2}    \pm 4.443 \times 10^{-2}$
  & $\hm 1.813 \times 10^{0\hm}    \pm 2.385 \times 10^{0}$ \\
  & \multicolumn{1}{r}{(107  \%)}
  & \multicolumn{1}{r}{(2.003  \%)}
  & \multicolumn{1}{r}{(44.8  \%)}
  & \multicolumn{1}{r}{(131.6  \%)} \\
\bottomrule
\bottomrule
\end{tabular}
\end{table*}

\begin{table*}[!t]
\caption{
    \label{table5}
Same as Table~\ref{table4}, but for $T = 500\C$.}
\centering
\setlength{\tabcolsep}{3pt}
\squeezetable
\begin{tabular}{l l l l l}
\toprule
\toprule
  \multicolumn{1}{c}{$k$}
  & \multicolumn{1}{c}{$a_k$}
  & \multicolumn{1}{c}{$\omega_k~[E_h/\hbar]$}
  & \multicolumn{1}{c}{$\gamma_k~[E_h/\hbar]$}
  & \multicolumn{1}{c}{$\gamma'_k~[E_h/\hbar]$} \\
\midrule
1 & $    5.687 \times 10^{0\hm}  \pm 1.616 \times 10^{-2}$
  & $    1.088 \times 10^{-3}    \pm 5.724 \times 10^{-7}$
  & $    2.837 \times 10^{-4}    \pm 1.025 \times 10^{-6}$
  & $\hm 5.422 \times 10^{-5}    \pm 4.019 \times 10^{-6}$ \\
  & \multicolumn{1}{r}{(0.2841  \%)}
  & \multicolumn{1}{r}{(0.0526  \%)}
  & \multicolumn{1}{r}{(0.3612  \%)}
  & \multicolumn{1}{r}{(7.413  \%)} \\[2.1ex]
2 & $    6.332 \times 10^{-2}    \pm 1.818 \times 10^{-2}$
  & $    4.040 \times 10^{-1}    \pm 2.012 \times 10^{-3}$
  & $    1.918 \times 10^{-2}    \pm 3.330 \times 10^{-3}$
  & $\hm 3.259 \times 10^{-1}    \pm 1.518 \times 10^{-1}$ \\
  & \multicolumn{1}{r}{(28.71  \%)}
  & \multicolumn{1}{r}{(0.4979  \%)}
  & \multicolumn{1}{r}{(17.36  \%)}
  & \multicolumn{1}{r}{(46.56  \%)} \\[2.1ex]
3 & $    1.264 \times 10^{-1}    \pm 2.908 \times 10^{-2}$
  & $    4.737 \times 10^{-1}    \pm 2.550 \times 10^{-3}$
  & $    2.835 \times 10^{-2}    \pm 4.507 \times 10^{-3}$
  & $   -9.653 \times 10^{-2}    \pm 1.197 \times 10^{-1}$ \\
  & \multicolumn{1}{r}{(23.01  \%)}
  & \multicolumn{1}{r}{(0.5383  \%)}
  & \multicolumn{1}{r}{(15.9  \%)}
  & \multicolumn{1}{r}{(124  \%)} \\[2.1ex]
4 & $    4.704 \times 10^{-1}    \pm 8.543 \times 10^{-2}$
  & $    5.321 \times 10^{-1}    \pm 6.257 \times 10^{-3}$
  & $    1.003 \times 10^{-1}    \pm 1.375 \times 10^{-2}$
  & $\hm 2.151 \times 10^{-1}    \pm 8.345 \times 10^{-2}$ \\
  & \multicolumn{1}{r}{(18.16  \%)}
  & \multicolumn{1}{r}{(1.176  \%)}
  & \multicolumn{1}{r}{(13.71  \%)}
  & \multicolumn{1}{r}{(38.79  \%)} \\[2.1ex]
5 & $    3.393 \times 10^{-1}    \pm 1.068 \times 10^{-1}$
  & $    1.043 \times 10^{0\hm}  \pm 7.818 \times 10^{-2}$
  & $    4.762 \times 10^{-1}    \pm 1.817 \times 10^{-1}$
  & $   -4.610 \times 10^{-1}    \pm 3.665 \times 10^{-1}$ \\
  & \multicolumn{1}{r}{(31.47  \%)}
  & \multicolumn{1}{r}{(7.496  \%)}
  & \multicolumn{1}{r}{(38.17  \%)}
  & \multicolumn{1}{r}{(79.51  \%)} \\[2.1ex]
6 & $    4.173 \times 10^{-2}    \pm 4.604 \times 10^{-2}$
  & $    1.234 \times 10^{0\hm}  \pm 2.546 \times 10^{-2}$
  & $    9.917 \times 10^{-2}    \pm 4.578 \times 10^{-2}$
  & $\hm 1.813 \times 10^{0\hm}  \pm 2.458 \times 10^{0}$ \\
  & \multicolumn{1}{r}{(110.3  \%)}
  & \multicolumn{1}{r}{(2.063  \%)}
  & \multicolumn{1}{r}{(46.16  \%)}
  & \multicolumn{1}{r}{(135.6  \%)} \\
\bottomrule
\bottomrule
\end{tabular}
\end{table*}

In the fitting, we have been
careful to avoid introducing a conceivable, numerically small
spurious negative imaginary part of the dielectric
function, {\em i.e.}, to implement the
condition $\Im[\epsilon(\omega)] \geq 0$,
especially in the transition region between
the IR and UV peaks, where the imaginary (absorptive)
part of the dielectric function assumes
very small numerical values
(see also Ref.~\cite{MoEtAl2025erratum}).
The implementation of this requirement is
relatively straightforward and can be
accomplished by introducing additional constraints in the
nonlinear fits~\cite{Wo1999}.

%
%
\subsection{Results}
\label{sec2D}

The fitting procedures outlined in the
previous sections lead to a satisfactory
representation of the optical data
for \CaFT{} in the temperature
range $22\C < T < 500\C$.
This corresponds to the range
$0 < T_\Delta < 1.71$ for the reduced
temperature.
Following the validation of the extraction procedure
for the UV data (see Fig.~\ref{fig1}),
we refer to Figs.~\ref{fig2} and~\ref{fig3} for
the overall characteristics of the
dielectric functions (real and imaginary parts)
in the range $0 < \omega < 2.2 \, {\rm a.u.}$,
and $0 < T_\Delta < 1.71$.
The real part of the dielectric function is
investigated in Fig.~\ref{fig4} for four
specific temperatures ($T = 22\,\C$,
$T = 100\,\C$, $T = 300\,\C$, $T = 500\,\C$).
The same is done in Fig.~\ref{fig5} for the
imaginary part.

From Fig.~\ref{fig6}, it becomes clear that our fit leads to a
satisfactory description of an excitonic (triple-)peak
structure found in Re$[\epsilon(\omega)]$
near photon energies of about 10.8\,eV.
This feature is essential for the understanding
of the excitations inside the crystal,
as further elaborated in Sec.~\ref{sec3}.
Finally, let us remark that
we shall comment on the statistical uncertainty
assigned to the UV fit parameters in Sec.~\ref{sec4A}
(see Tables~\ref{table4} and~\ref{table5},
and Appendix~\ref{appa}).

%
%
\section{First--Principles Calculations}
\label{sec3}

In this section, we discuss the zero-temperature optical spectra of CaF$_2$
obtained using TDDFT and the random
phase approximation (RPA) \cite{OnReRu2002,Ul2012,UlYa2015} to compare with the
experimental data. We mention that there exist several earlier calculations of
the optical spectra of CaF$_2$ \cite{GaEtAl1992,BeSh1999,MaRo2007}; however,
these studies do not cover the broad energy range of interest here (up to 50
eV) and generally do not agree too well with the available experimental data.

The Kohn--Sham band structures of fcc CaF$_2$ were obtained using the
Perdew--Burke--Ernzerhof (PBE)
functional~\cite{PeBuEr1996} within the {\sc Quantum Espresso}
package~\cite{QUANTUM_ESPRESSO}, implementing a plane-wave basis and optimized
norm-conserving Vanderbilt pseudopotentials \cite{Ha2013}.  We used the
experimental lattice constant $10.424$\,a.u. \cite{HaFi1981}
and a $16 \times 16 \times 16$ $k$-point grid.

\begin{figure}[t!]
\begin{center}
\includegraphics[width=0.91\linewidth]{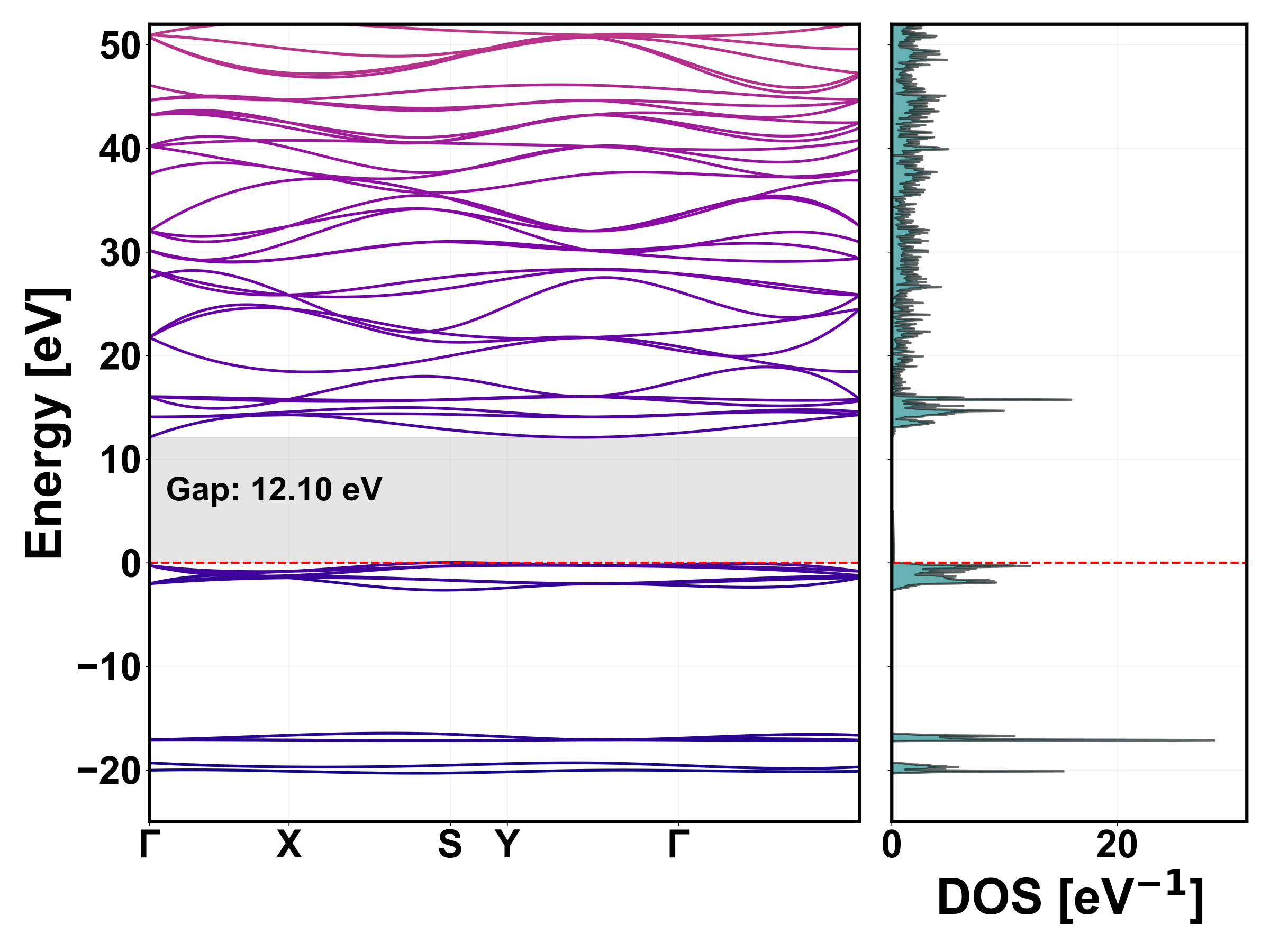}
\caption{ \label{fig7}
Electronic band structure and total density of states
(DOS) of CaF$_2$, obtained using the PBE exchange-correlation functional,
adjusting the band gap to its experimental value of
$12.10$\,eV~\cite{Ru1972,ChZhGuRo2022}. 
The red dashed line indicates the Fermi level at $0$ eV. }
\end{center}
\end{figure}

The resulting band structure and electronic density of states (DOS) are shown
in Fig. \ref{fig7}. The energy dispersion was computed along the high-symmetry
path $\Gamma-X-S-Y-\Gamma$ in the first Brillouin zone, revealing a direct band
gap which was adjusted to the experimental value of 
$12.10$\,eV~\cite{Ru1972,ChZhGuRo2022} using scissors
operators \cite{LeAl1989}, setting the Fermi level as the
reference energy (0 eV). The strongly
insulating nature of CaF$_2$ is reflected in the flat
dispersions of the occupied bands. The 6 bands right below the valence band
edge originate from the $2p$ electrons of F; the lower lying valence bands come
from the $3s$ and $2p$ electrons of Ca and the $2s$ electrons of~F.
The lowest conduction bands originate from the
$3s$ and $3d$ levels of Ca and the $3s$ levels of F.

\begin{figure}[t!]
\begin{center}
\includegraphics[width=1.0\linewidth]{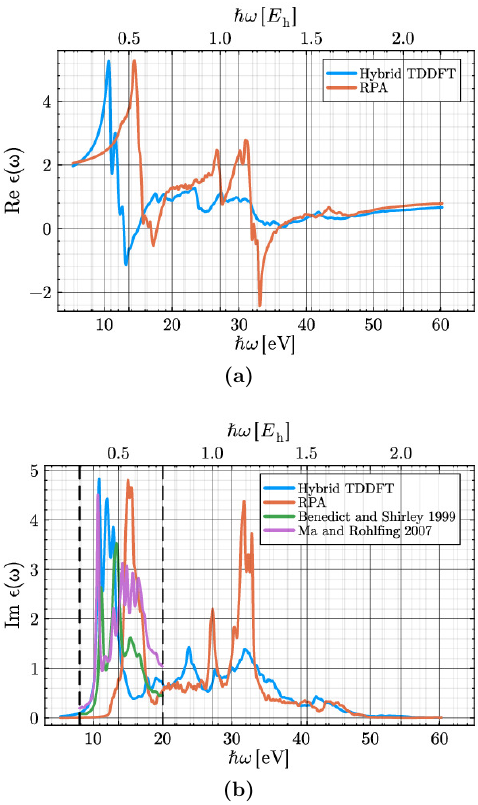}
\caption{ \label{fig8}
Comparison of the real and imaginary parts,
Re$[\epsilon(\omega)]$ [panel(a)] and Im$[\epsilon(\omega)]$
[panel (b)], of the macroscopic dielectric
function of CaF$_2$, calculated using the hybrid-TDDFT approach and the RPA.
The excitonic enhancement of the spectrum below the gap at 10.8\,eV is clearly
visible in the hybrid-TDDFT spectra. 
For the imaginary part, we supplement the figure with a green 
curve (theoretical predictions reported by Benedict and Shirley 
in Ref.~\cite{BeSh1999}) 
and a purple curve
(theoretical predictions reported by Ma and Rohlfing in Ref.~\cite{MaRo2007}).
The energy range investigated in Refs.~\cite{BeSh1999,MaRo2007}
($8\,\mathrm{eV} < \hbar \omega <  20\,\mathrm{eV}$)
is indicated by vertical dashes.}
\end{center}
\end{figure}

To investigate the optical response of CaF$_2$
we calculate the macroscopic dielectric function $\epsilon(\omega)$  with
the {\sc Yambo} code \cite{MaHoGrVa2009}, using a customized implementation of
TDDFT with a dielectrically screened hybrid functional
\cite{SuYaUl2020,SuUl2021,AlSuUl2025}. Hybrid functionals are constructed by
combining a fraction of nonlocal (Hartree--Fock) exchange with semilocal density
functional approximations; here, the mixing parameter is determined via the
inverse static dielectric constant of the material.  As shown in previous work
\cite{SuYaUl2020,SuUl2021,AlSuUl2025}, dielectrically screened hybrid
functionals are essentially as accurate as the Bethe-Salpeter equation (which
is generally considered as most reliable method
for calculating optical properties of materials~\cite{OnReRu2002}),
but at a fraction of the computational cost. In particular,
excitonic effects in optical spectra are reproduced with quantitative accuracy;
by contrast, random-phase-approximation (RPA)
spectra fail to account for excitons, although otherwise they
often tend to capture the optical response qualitatively correctly.

In Fig.~\ref{fig8}, we present the real and imaginary parts of the dielectric
function $\epsilon(\omega)$ of CaF$_2$, calculated using the RPA and the
hybrid-TDDFT approach. The calculations were done using 10 valence bands and 28
conduction bands as input for the linear response calculation in {\sc Yambo},
which covers the spectral range  5 eV $ < \hbar \omega < $ 50 eV.  The number
of bands was carefully checked to ensure convergence of the spectra; we also
notice that the spectra were calculated with a numerical broadening of
$0.3$\,eV to simulate experimental linewidths.  In Ref.~\cite{LeStRo1965}, the
experimental data were taken for cleaved single-crystalline samples. The
standard cleavage plane for crystals with a face-centered cubic (fcc) lattice
is the (111) plane (see also Appendix~\ref{appb}). We thus calculate the
optical response along the [111] crystallographic direction; calculations along
the [100] direction give identical results, consistent with the fact that the
optical response for cubic crystals is isotropic.

The RPA spectra, which can be viewed as the independent-particle response of
the material, are obtained without any dynamical many-body effects;
this means, in particular, that RPA does not include
electron-hole interactions and hence does not produce any excitons. The first
prominent RPA peak of Im$[\epsilon(\omega)]$ at
15\,eV can be directly related to
the DOS shown in Fig.~\ref{fig7}: it is caused by transitions between the
highest occupied valence bands (which produces a sharp peak at the Fermi level)
and the fourth and fifth conduction band (which produce a strong and very sharp
peak at 15 eV). RPA transitions to the lower conduction band edge have a much
smaller oscillator strength, giving rise only to a small shoulder around
12.5\,eV.

\begin{figure}[t!]
\begin{center}
\includegraphics[width=1.0\linewidth]{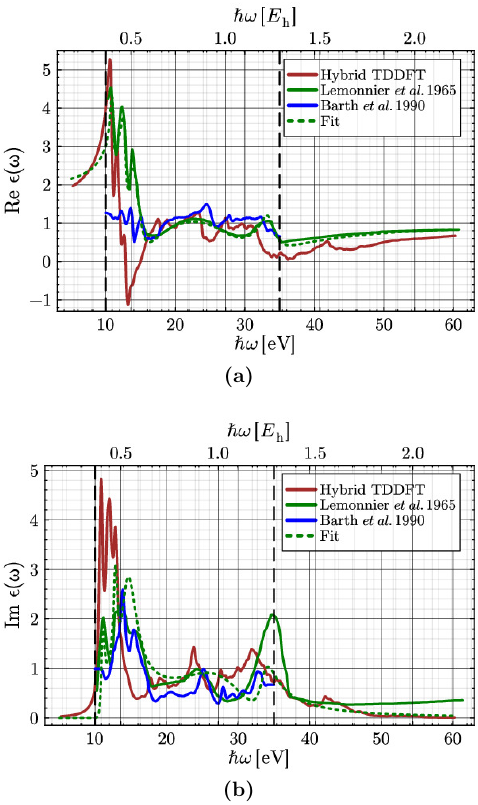}
\caption{ \label{fig9}
Comparison of the real
and imaginary parts [see panels (a) and (b), respectively]
of the macroscopic dielectric function of CaF$_2$,
calculated using the hybrid-TDDFT approach
with the experimental data and with our fit. 
While our fit is based on the experimental data
from Ref.~\cite{LeStRo1965} (green curves), 
we include, for comparison, additional experimental data from
Barth {\em et al.}~(Ref.~\cite{BaEtAl1990}, blue curves). 
The excitonic peak at about 10.8 eV is close
to the lower end of the experimental 
photon energy range investigated in Ref.~\cite{BaEtAl1990}
($10\,\mathrm{eV} < \hbar \omega < 35\,\mathrm{eV}$,
marked by the vertical dashes).
As noted by Benedict and Shirley on p.~5449 of Ref.~\cite{BeSh1999},
the excitonic peak is not resolved
in the data reported in Ref.~\cite{BaEtAl1990}.
A detailed resolution of the discrepancies between the data
sets communicated in Refs.~\cite{LeStRo1965,BaEtAl1990},
over the entire UV frequency region,
lies beyond the scope of the present work.}
\end{center}
\end{figure}

The hybrid-TDDFT spectrum, by contrast, agrees quite
well with the experimental
data as shown in Fig.~\ref{fig9}, exhibiting a very strong, three-fold split peak of
Im$[\epsilon(\omega)]$ in the energy range between 10 and 15 eV, due to excitonic
interactions. The first excitonic peak right at the band gap is more prominent
in the calculation compared to experiment.  For Re$[\epsilon(\omega)]$, the
experimental data show a very pronounced three-fold splitting, which is also
seen in the TDDFT-calculated curve,
although the third peak only appears as a small shoulder.

At higher energies, we observe that the broad experimental signals at 
25\,eV and 30\,eV are also found in the 
hybrid-TDDFT spectra (the corresponding RPA features
are found to be blueshifted by a couple of eV). The relative peak heights are
roughly equal in the calculated spectra, whereas the experimental spectra show a
much stronger peak at 30\,eV. These differences, as well as the minor
discrepancies in the excitonic part of the spectra discussed in the preceding
paragraph, are most likely due mainly to the approximations used for the input
band structure, which was calculated using the PBE functional plus scissors.
For possible future studies, we remark that band
structures obtained with (more computationally involved)
many-body techniques such as $GW$~\cite{OnReRu2002}
may lead to an even better agreement between theoretical and
experimental band structures.

\begin{figure}[t!]
\begin{center}
\includegraphics[width=0.91\linewidth]{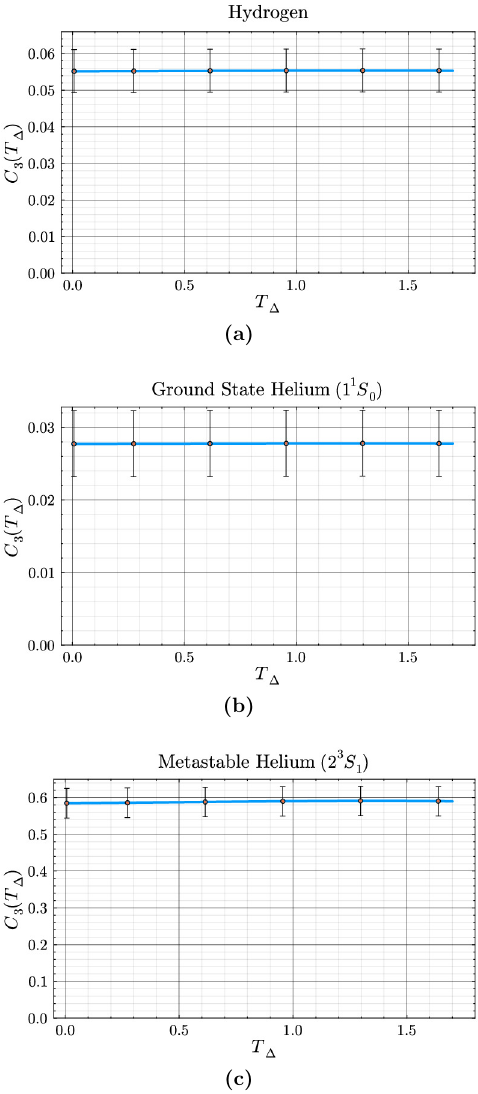}
\caption{
\label{fig10} $C_3(T_\Delta)$ coefficients
for hydrogen [panel (a)],
ground-state helium ($1\,{}^1S_0$) [panel (b)] and
metastable helium ($2\,{}^3S_1$) [panel (c)].}
\end{center}
\end{figure}

\begin{figure}[t!]
\begin{center}
\includegraphics[width=0.91\linewidth]{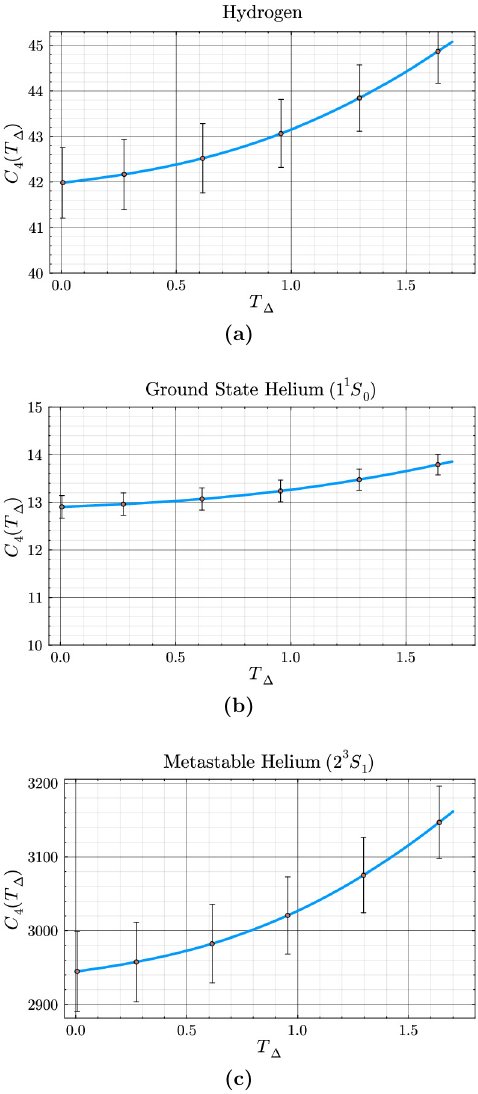}
\caption{\label{fig11}
$C_4(T_\Delta)$ coefficients for
hydrogen [panel (a)], helium
[panel (b)], and metastable helium
[panel (c)] interacting with \CaFT{}.}
\end{center}
\end{figure}

\begin{table*}
\begin{center}
\begin{minipage}{0.77\linewidth}
\caption{
\label{table6}
Temperature-dependent short--range, $C_3(T_\Delta)$ and long--range
$C_4(T_\Delta)$ coefficients for hydrogen, helium, and metastable helium
interacting with \CaFT. Results are given in atomic units
[see also the discussion following Eq.~\eqref{C3}].
}
\renewcommand{\arraystretch}{1.5}
\setlength{\tabcolsep}{8pt}
\begin{tabular}{@{}c@{\hskip 10pt}ccccc@{\hskip 6pt}c@{}}
\toprule
\toprule
& \multicolumn{2}{c}{Hydrogen} &
\multicolumn{2}{c}{Helium} &
\multicolumn{2}{c}{Metastable Helium\hphantom{\;\,}} \\
\cmidrule(r){2-3}
\cmidrule(r){4-5}
\cmidrule(r){6-7}
$T\,[\C]$ & $C_3$ & $C_4$ & $C_3$ & $C_4$ & $C_3$ & $C_4$ \\
\midrule
22  & 0.05522 & 42.003 & 0.02773 & 12.911 & 0.5848 & 2946.111 \\
100 & 0.05523 & 42.133 & 0.02774 & 12.951 & 0.5851 & 2955.209 \\
200 & 0.05533 & 42.498 & 0.02778 & 13.063 & 0.5884 & 2980.797 \\
300 & 0.05539 & 43.158 & 0.02780 & 13.266 & 0.5907 & 3027.107 \\
400 & 0.05536 & 43.769 & 0.02779 & 13.453 & 0.5899 & 3069.935 \\
500 & 0.05537 & 44.886 & 0.02779 & 13.797 & 0.5905 & 3148.273 \\
\bottomrule
\bottomrule
\end{tabular}
\end{minipage}
\end{center}
\end{table*}

\begin{figure}[t!]
\includegraphics[width=0.8\linewidth]{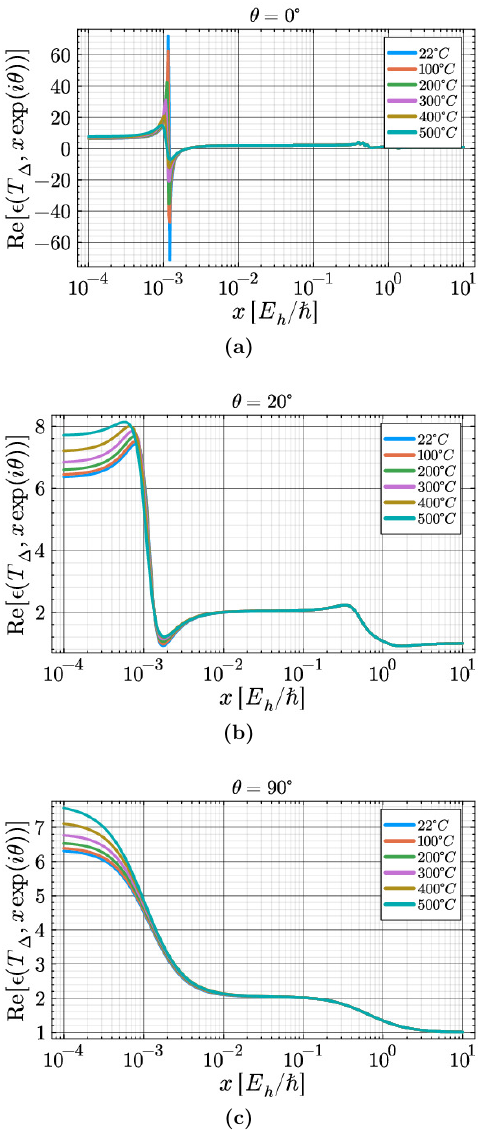}
\caption{\label{fig12}
Temperature dependence of the dielectric function $\epsilon(\omega)$
under Wick rotation, where $\omega = x \exp(\ii\theta)$. As the rotation angle
$\theta$ increases from $\theta=0^\circ$ to $\theta=90^\circ$,
the temperature--dependence of $\epsilon(\omega)$ rapidly diminishes.  Even at
$\theta=30^\circ$, the influence of temperature is already
substantially suppressed.  }
\end{figure}

\begin{figure}[t!]
\begin{center}
\includegraphics[width=0.88\linewidth]{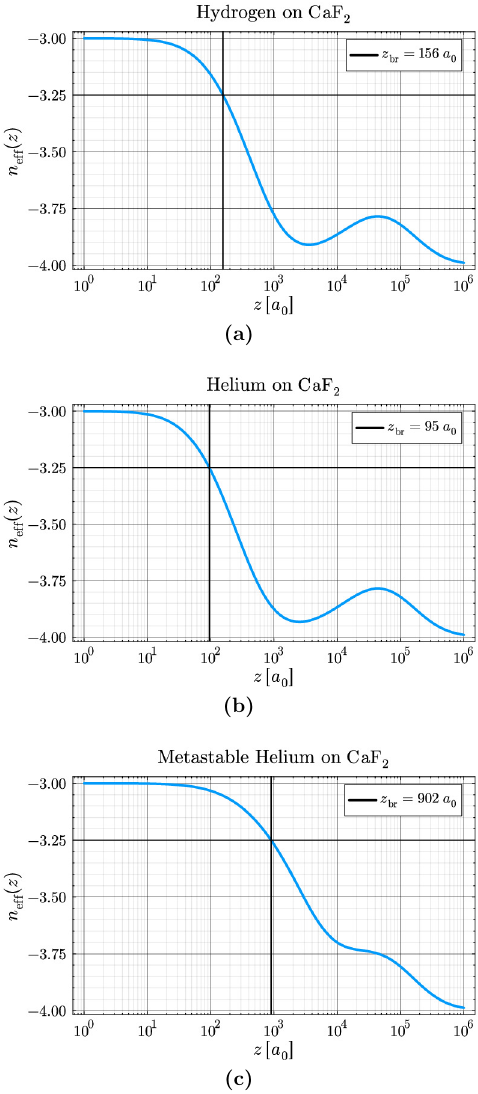}
\caption{
\label{fig13}
Effective exponent $\neff$ as a function of the atom-wall separation,
$z$, for hydrogen, ground-state helium, and metastable helium ($2\,{}^3S_1$)
interacting with \CaF2.  The potential is evaluated numerically, and $\neff$ is
calculated using Eq.~(\ref{neff}).  The dielectric function of \CaF2 is evaluated
at room temperature. The small but interesting hump of the effective exponent
$\neff$ toward a partial restoration of the nonretarded functional dependence
in between $10^4$ and $10^5$ atomic units is due to the
presence of the giant IR resonance of \CaF2. See text for further explanations.
}
\end{center}
\end{figure}

%
%
\section{Atom--Surface Potentials}
\label{sec4}

%
%
\subsection{Short--Range Limit}
\label{sec4A}

It is well known (see Appendix A of Ref.~\cite{DaUlJe2024}
and references therein) that the Casimir interaction
between an electrically neutral atom and a dielectric
surface is approximated by
a $1/z^3$ formula for short range,
which transitions into a
$1/z^4$ dependence for long range,
before transitioning into an
extreme-long-range, $1/z^3$ regime,
(for distances above the thermal wavelength),
where the interaction energy
between the atom and the solid
is numerically small and proportional
to the absolute temperature.
Suitable functional forms for the
atom-surface interactions in intermediate
regions have been discussed in Ref.~\cite{JeMo2023blog}.
In the short-range limit, the atom-surface potential $V(z)$, with $z$ denoting
the atom-surface distance can be expressed as
\begin{equation}
\label{C3}
V(z) \approx - \frac{C_3}{z^3}, \quad
a_0 \sim d \ll z \ll z_{\rm crit} \,.
\end{equation}
Here, $a_0$ is the Bohr radius,
$\alpha$ is the fine structure constant,
and $d$ is a characteristic
distance scale of the crystal lattice.
Our use of atomic units implies that,
if $z$ is measured in units of the Bohr radius,
then $V(z)$ is obtained in units of Hartrees.
Further clarifications can be found in Eq.~(2) of
Ref.~\cite{CrGuRe2019} as well as Refs.~\cite{Kl2025comment,Je2025reply}.
Via the definition of a reference plane $z_0$, the validity of the
short-distance expansion can be extended to the
physisorption range~\cite{ZaKo1976,Je2025reply}
(see also Appendix~\ref{appb}).
The reference-plane position $z_0$ is
roughly equal to the center of the induced charge density
inside the solid (see Ref.~\cite{ZaKo1976},
Chap.~2 of Ref.~\cite{BrCoZa1997},
and Chap.~6 of Ref.~\cite{Li1997}).
For the physisorption range $z \gtrsim a_0$,
one replaces $1/z^3$ in Eq.~\eqref{C3} by $1/(z - z_0)^3$
(see Refs.~\cite{ZaKo1976,Je2025reply}).

The critical distance $z_{\rm crit}$,
which marks the upper limit of the applicability
of the short-range expansion, is given by~\cite{DaUlJe2024}
\begin{equation}
\label{zcrit}
z_{\rm crit} = \frac{a_0}{\alpha}
\sqrt{ \frac{\alpha_{\rm a.u.}(0)}{Z} } \,,
\end{equation}
where $\alpha_{\rm a.u.}(0)$ is the
static polarizability of the atom,
measured in atomic units.
(Of course, strictly speaking,
we use atomic units throughout this article,
so that the indication of $a_0$,
which is unity in atomic units,
could be considered as redundant.
However, we still indicate $a_0$
in some of our formulas for clarity.)

In atomic units, the $C_3$ coefficient can be written as
\begin{equation}
\label{epsilon}
C_3 = \frac{1}{4\pi} \int_0^\infty \dd\omega~\alpha_{\rm a.u.}(\ii\omega)
\frac{\epsilon(\ii\omega) - 1}{\epsilon(\ii\omega) + 1},
\end{equation}
where $\alpha(\ii\omega)$ is the dynamic polarizability
of the atom, evaluated at imaginary frequency argument,
and $\epsilon(\ii\omega)$ denotes the
dielectric function, evaluated likewise at imaginary frequencies.
In the following, we consider the
evaluation of the $C_3$ coefficient for interactions
with a \CaFT{} surface, where the
dielectric function $\epsilon(\ii \omega)$,
for imaginary argument $\ii \omega$ with $\omega \in (0, \infty)$,
is obtained according to Eq.~\eqref{master}.

The $C_3$ coefficients for hydrogen, ground state helium and metastable helium
($2\,{}^3S_1$) interacting with \CaFT{} are shown in Fig.~\ref{fig10}.
Helium is one of the preferred atoms for
studies of quantum reflection occurring near
various crystal surfaces~\cite{NaEdMa1982,DoEtAl1993,DKDuHaLa2001,DrDK2003},
and planar \CaFT{} surfaces provide for a suitable substrate
in quantum reflection studies~\cite{DKpriv}.
Hydrogen (and its antiparticle) have also been advocated as
sensitive probes for quantum reflection
studies above surfaces~\cite{CoSeRa1998,DuEtAl2013}.
For the IR and the visible range,
we conservatively assume a $5\%$ uncertainty in our
dielectric function model, as explained in Sec.~\ref{sec2C}.
The conservative estimate
of a 5\,\% uncertainty takes care of conceivable unaccounted
systematic effects in the ellipsometry
data~\cite{PSEtAl2009,DaMa2002,FiSaEnSu2008,%
LeEtAl2015temp,KGEtAl2017,ZhWaTh2023},
and is in line with typical accuracy limits of ellipsometry
measurements for comparable materials~\cite{MoEtAl2022,MoEtAl2025erratum}.
The wavelength range $184\,{\rm nm} < \lambda < 2326\,{\rm nm}$
is covered by available data sheets from
industrial applications~\cite{CaF2corning,CaF2hellma},
and by Ref.~\cite{ZhWaTh2023} (for completeness,
we indicate the corresponding frequency range
in atomic units, $0.01959 \, {\rm a.u.} <
\omega < 0.2476 \, {\rm a.u.}$).
In view of excellent
mutual agreement between Refs.~\cite{CaF2corning,CaF2hellma,ZhWaTh2023}
and our fit in the indicated wavelength range
$184\,{\rm nm} < \lambda < 2326\,{\rm nm}$,
we assume a $1\%$ uncertainty in our
dielectric function model in this range.
For the UV range, we conclude that either an unknown systematic effect
in the experiment or the approximate character of
our first-principles calculations
must be responsible for the discrepancies between
theory and experiment as reported
in Figs.~\ref{fig6} and~\ref{fig9}.  These discrepancies are
also manifest in the comparatively large
statistical uncertainties of the fit parameters
listed in Tables~\ref{table4} and~\ref{table5}
(for details of the statistical analysis,
see Appendix~\ref{appa}). We have checked that the UV region has a
negligible influence on the calculation of the $C_3$
coefficient. Hence, the $5\%$ uncertainty in our
dielectric function model, in the relevant
frequency range, translates into a 5\,\%
uncertainty in our $C_3$ coefficients.
For the dynamic polarizability of hydrogen,
$\alpha(\ii\omega)$, we use the analytic expression in Eq.~(4.161) of
Ref.~\cite{JeAd2022book}. For ground-state and metastable helium, the
polarizability is calculated using fully correlated basis sets of exponential
basis functions, as outlined in Ref.~\cite{DaUlJe2024}.
Numerical values are presented in Table~\ref{table6}.

The surprisingly smooth dependence of the $C_3$
coefficient on temperature, as evident from Fig.~\ref{fig10},
despite the rather drastic dependence of the
IR peak on the temperature (see Figs.~\ref{fig2} and~\ref{fig3}),
finds an explanation in the remarkable
reduction of the temperature dependence
after the rotation of the $\omega$ integration
contour into the complex plane
(see also Figs.~\ref{fig11} and~\ref{fig12}).

%
%
\subsection{Long--Range Interaction}
\label{sec4B}

In the distance range
\begin{equation}
z_{\rm crit} \ll z \ll \lambda_T = \frac{\hbar c}{k_B T} \,,
\end{equation}
where $\lambda_T$ is the thermal wavelength and
$k_B$ is the Boltzmann constant,
the retarded (Casimir-Polder) regime of atom-surface
interaction potential, $V(z)$, can be approximated
(see Appendix~A of Ref.~\cite{DaUlJe2024}) by
the formula
\begin{equation}
\label{C4}
V(z) \approx - \frac{C_4}{z^4},
\quad C_4 = \frac{3}{16 \pi\alpha}~\alpha(0) \,
\int_1^\infty \dd p \frac{H(\epsilon(0),p)}{p^4} \,.
\end{equation}
The function $H \equiv H(\epsilon, p)$ is given by
(see Ref.~\cite{Je2024multipole} and references therein)
\begin{equation}
\label{defH}
H(\epsilon,p) \!\! = \!\!\frac{\sqrt{\epsilon-1+p^2}-p}{\sqrt{\epsilon-1+p^2}+p}
+ (1-2p^2) \frac{\sqrt{\epsilon-1+p^2}-p \epsilon}%
{\sqrt{\epsilon-1+p^2}+p \epsilon}.
\end{equation}
The $C_4$ coefficients for hydrogen, ground state helium and metastable helium
($2\,{}^3S_1$) interacting with \CaFT{} are shown in Fig.~\ref{fig11}.

\subsection{Intermediate Region: Onset of Retardation}

The transition from the short-range (van-der-Waals) to the long-range
(Casimir-Polder) asymptotics, due to the retardation effect, has been discussed
in detail in Ref.~\cite{DaUlJe2024}.  The intermediate region between the two
asymptotics can be conveniently
studied using an effective ``local'' exponent $\neff$, defined
in terms of the logarithmic derivative of the interaction potential $V(z)$,
\begin{equation}
\label{neff}
n_{\rm eff}(z)
= \frac{z}{V(z)} \frac{\dd V(z)}{\dd z}
= \frac{\dd \ln(|V(z)|)}{\dd \ln z},
\end{equation}
where the logarithm of the potential refers to the logarithm of its numerical
value, expressed in atomic units.  For a potential $V(z) = V_0 z^n$, the
effective exponent, $\neff$, evaluates to exactly $n$, that is,
one would obtain
$n_{\rm eff} = -3$ for a potential of the functional
form $V(z) = -C_3/z^3$,
while $n_{\rm eff} = -4$ for a potential of the functional
form $V(z) = -C_3/z^4$.

We have used~\cite{DaUlJe2024} the value of $\neff = -3.25$
to mark the distance $z_{\rm br}$ where the short-range expansion breaks
down. The dependence of $\neff$ on the atom-wall distance
is shown in Fig.~\ref{fig13}, for
hydrogen, ground-state helium, and metastable ($2\,{}^3S_1$) helium
interacting with \CaF2 at room temperature.
We observe a small but interesting hump of the effective exponent
$\neff$ toward a partial restoration of the nonretarded functional dependence
in between $10^4$ and $10^5$ atomic units. This hump is due to the
presence of the giant IR resonance of \CaF2.
We recall that retardation sets in
(see Chap.~5 of Ref.~\cite{JeAd2022book}) when the virtual photons,
which mediate the atom-surface interaction, acquire a
phase commensurate with unity during their travel
to and from the surface. This is the case when the largest characteristic
frequency $\omega_{\rm ch}$ of the combined atom-solid system,
the atom-wall distance $z$, and the speed of light $c$ fulfill
the condition $\omega_{\rm ch} \, z / c \approx 1$.
In typical cases, $\omega_{\rm ch}$ is determined by the
atom's characteristic frequency, and this observation leads to
the estimate given in Eq.~\eqref{zcrit}. For \CaF2, the giant IR resonance
is so strong that, between $10^4$ and $10^5$ atomic units, the
effective exponent $\neff$ displays a (very partial) restoration
toward the nonretarded interaction. However, even in the
distance region between $10^4$ and $10^5$ atomic units, the
restoration is only partial, and the effective exponent
does not rise above $-3.6$, near the second hump,
for all systems under study here
(see Fig.~\ref{fig13}). In Table~\ref{table7}, we compare
the breakdown distance, $z_\brek$, with the critical distance, $z_\crit$
(defined in Eq.~\ref{zcrit}).
We also include the $z_\brek$ values for the
same atoms interacting with silicon and gold, as presented in
Ref.~\cite{DaUlJe2024}. A good order-of-magnitude agreement
between $z_\crit$ and $z_\brek$ is found, just like in
Ref.~\cite{DaUlJe2024} for silicon and gold.

Let us now discuss the location of the
bump in Fig.~\ref{fig13} in detail.
To this end, we first have to remember the
interpolating formula between Eqs.~\eqref{C3}
and~\eqref{C4}, which has been given in
in Eqs.~(18) and~(21) of Ref.~\cite{AnPiSt2004},
and in Eqs.~(2) and~(3) of Ref.~\cite{DaUlJe2024}.
It reads as follows,
\begin{align}
\label{Vatom}
V(z) =& 
-\frac{\alpha^3}{2 \pi} \int\limits_0^\infty \dd\omega\,\omega^3 \,
\alpha(\ii \omega)
\int\limits_1^\infty\dd p \,
\ee^{- 2 \, \alpha\, p  \, \omega \, z }
H( \epsilon( \ii \omega), p )\, ,
\end{align}
where $H(\epsilon, p)$ is given in Eq.~\eqref{defH}.
The transition to the long-range regime
is determined by the exponential factor
$\exp\left( - 2 \, \alpha\, p  \, \omega \, z  \right)$.
The nonretarded regime is obtained when,
for $p \geq 1$, the argument of the exponential
is small against unity for $\omega \approx \omega_{\rm ch}$,
where $\omega_{\rm ch}$ is a characteristic
frequency of the system.
By contrast, the retarded regime is obtained when,
for $p \geq 1$, the argument of the exponential
is large against unity for $\omega \approx \omega_{\rm ch}$.
The transition to the retarded regime
occurs when the argument of the exponential is
of order unity,
\begin{equation}
2 \, \alpha\, \omega_{\rm ch} \, z_\crit = 1 \,,
\qquad
z_\crit = \frac{ 1 } { 2 \, \alpha\, \omega_{\rm ch} } \, .
\end{equation}
For distances $z \geq z_\crit$, retardation sets in.
As it was stressed in Ref.~\cite{DaUlJe2024}, it is thus
the {\em largest} characteristic frequency $\omega_{\rm ch}$ in
the problem which
determines the onset of retardation.
In many typical cases, this is a characteristic frequency of the
{\em atom}, not of the {\em surface} material~\cite{DaUlJe2024}.
For the giant IR resonance of \CaFT{}, on the other hand,
a second (numerically less
significant) transition region is determined
by the location of the IR resonance, which, according to
Figs.~\ref{fig4} and~\ref{fig5}, lies at
$\omega'_{\rm ch} \approx 0.0014 \, {\rm a.u.}$.
The corresponding critical distance is
(in atomic units)
\begin{equation}
\label{zpcrit}
z'_\crit = \frac{ 1 } { 2 \, \alpha\, \omega'_{\rm ch} }
\approx 49\,000 \, {\rm a.u.}\,,
\end{equation}
in excellent agreement with the results shown in
Fig.~\ref{fig13}.

\begin{table}
\caption{
\label{table7}
Comparison of the numerically
calculated breakdown distance, $z_\brek$, with the
analytically estimated critical
distance, $z_\crit$, for hydrogen, helium and metastable helium
interacting with \CaF2, silicon, and gold.
The dielectric function of the surfaces is
calculated for room temperature.
The validity of the order-of-magnitude estimate
$z_\crit \sim z_\brek$ is confirmed.}
\setlength{\tabcolsep}{8pt}
\begin{tabular}{lllll}
\toprule
\toprule
& $z_\crit~[a_0]$ & \multicolumn{3}{c}{$z_\brek~[a_0]$}  \\
\cmidrule(lr){3-5}
                   & (any solid) & \CaF2 & Si   & Au  \\
\midrule
Hydrogen              & 290      & 156   & 203  & 309   \\
Helium ($1 {}^1 S_0$) & 113      & 95    & 126  & 228   \\
Helium ($2 {}^3 S_1$) & 1720     & 902   & 979  & 1336  \\
\bottomrule
\bottomrule
\end{tabular}
\end{table}

%
%
\section{Conclusions}
\label{sec5}

Let us briefly review the most important
conclusions of the current paper, which are fourfold:
\begin{itemize}
\item[{\em (i)}] The dielectric function of \CaFT{}{}
can be described to very good accuracy
by the RRCO model given in Eq.~\eqref{master},
in a compact functional form, spanning wide photon
energy ranges from the static limit to the far UV 
($\hbar \omega \approx 50$\,eV).
Previous investigations were limited
to relatively narrow spectral ranges
({\em e.g.}, Ref.~\cite{PSEtAl2009} for the IR peak,
Ref.~\cite{ZhWaTh2023} for the transparent region,
and Ref.~\cite{LeStRo1965} for the UV).
Still, the authors of  Ref.~\cite{PSEtAl2009} 
relied on involved functional forms
for the analytic description of the 
frequency-dependent dielectric function~\cite{PSEtAl2009}
containing Dawson integrals;
our {\em ansatz} from Eq.~\eqref{master} 
covers a much wider frequency range and
is more compact in its analytic form.

\item[{\em (ii)}] We present {\em ab initio} calculations for the 
dielectric function in the UV range
which show a semi-quantitative agreement
with the data of Ref.~\cite{LeStRo1965},
confirming prospects of an improved
understanding of the dielectric function in the 
UV over wide frequency ranges, conceivably
on the basis of future experiments at modern 
synchrotron facilities~\cite{NeEtAl2014,ScEtAl2016}.

\item[{\em (iii)}] Our unified model,
spanning wide frequency ranges, enables the calculation
of atom-surface potentials for atoms in the 
vicinity of a calcium fluoride surface.
The corresponding integrals over virtual 
photons rely on the complete knowledge
of the dielectric function in the interval
$\hbar \omega \in (0, 50 \, {\rm eV})$,
which is provided here.

\item[{\em (iv)}] We uncover two unexpected 
phenomena in the atom-surface interactions.
First, the giant IR resonance of 
calcium fluoride leads to a partial ``revival''
of nonretardation at an atom-surface 
distance of about 49\,000 atomic units,
in full agreement with the characteristic 
photon frequency corresponding to the onset 
of retardation. Second, we find that, despite a
very pronounced temperature dependence of the 
giant IR peak of the dielectric function,
the temperature dependence of the 
$C_3$ and $C_4$ coefficients is not very pronounced.
Specifically, the $C_3$ coefficients given in
Table~\ref{table6} differ by less than 2\% over the temperature 
range $22^\circ \, {\rm C} < T < 500^\circ {\rm C}$,
while the peak of the real part of the 
dielectric function in the IR
varies by more than a factor four; namely, it
varies from $[\Re \; \epsilon]_{\rm max} \approx 80$ for 
$T = 22^\circ \, {\rm C}$ to a value of 
$[\Re \; \epsilon]_{\rm max} \approx 15$ for 
$T = 500^\circ \, {\rm C}$ (see Fig.~\ref{fig4}).
The explanation is provided by an unexpected damping of 
the temperature dependence upon the Wick 
rotation of the integration contour for the 
virtual photon (see Fig.~\ref{fig12}).
\end{itemize}

After giving the main results ``in a nutshell'',
we are now in the position to 
provide a further digression on particular aspects.
Our analysis, for the IR peak,
profit from the seminal paper~\cite{PSEtAl2009},
where the temperature-dependence of the IR peak was
analyzed in detail (and very involved functional forms were used in the
fits). We find that the RRCO model,
given in Eq.~\eqref{master}, provides for a convenient and simple
functional form. We have recently shown~\cite{DaUlJe2025coupled}
that the $\gamma'_k$ parameters can arise from the coupling
of different resonances. The IR resonance
constitutes a relatively isolated resonance which does not
significantly overlap with the UV peaks,
and, hence, the coupling to other resonances can be
assumed to be small (there could still be
radiative-reaction terms contributing to $\gamma'_1$).
As evident from Table~\ref{table1},
the $\gamma'_1$ parameter is significantly smaller
than $\gamma_1$ for all temperatures, consistent with the
observations reported in Ref.~\cite{DaUlJe2025coupled}.

For the UV peaks, by contrast, the $\gamma'_{k=2,3,4,5,6}$
parameters are larger than the corresponding $\gamma_k$
parameters (with the exception of $k=5$, where
the two parameters are nearly equal in magnitude).
These resonances correspond to lattice-modified
molecular transitions (interband transitions), and a considerable coupling
between them can be expected. TDDFT-based first-principles calculations
support our results, showing that the optical properties of CaF$_2$ in the region
at and below the band gap are dominated by excitonic effects.

One observation is interesting in this context. The pronounced
IR peak has the desired structure and does not create a
negative imaginary part ($\gamma_1 > \gamma'_1$ for
all temperatures). However, the UV peaks are intertwined
and do not have this property, {\em i.e.},
the overlapping resonances do not fulfill the
condition $\gamma_k > \gamma'_k$ separately.
The same behavior has been observed for
intrinsic silicon in Refs.~\cite{MoEtAl2022,MoEtAl2025erratum}.

The temperature-dependent universal
parameters of our model are given in
Tables~\ref{table2} and~\ref{table3}.
The functional form of the temperature-dependence
of the parameters of the IR peak
is given in Eq.~\eqref{IRtempFit}.
Our fit covers the range $0 < \omega < 2.2 \, {\rm a.u.}$ and
$0 < T_\Delta < 1.6$,
where $T_\Delta = (T - T_0)/T_0$,
and $T_0 = 293\,{\rm K}$ is the room temperature
[see Eq.~\eqref{defTdelta}].
The statistical quality of our fit is
discussed in Appendix~\ref{appa},
and an application to atom-surface
interactions is presented in Sec.~\ref{sec4}.
The latter illustrates that,
after a rotation of the argument of the
dielectric function into the complex plane,
$\omega \to \ii \omega$, the temperature-dependence
becomes smoother, and the
atom-surface interaction potentials
display only a marginal variation with the
temperature-dependent dielectric function.

The ``giant IR resonance'' is very much
temperature-dependent, and we here obtain
a simple parameterization.
The results obtained here for \CaFT{} are, in some sense,
more satisfactory than those obtained for intrinsic Si
in Refs.~\cite{MoEtAl2022,MoEtAl2025erratum},
because of the availability of experimental
data over a larger frequency range ($\hbar \omega \leq 60$\,eV
versus $\hbar \omega \leq 4.4 \; {\rm eV}$).
Furthermore, we here show that the RRCO model
is applicable in the case of a rather unusual
material (in terms of the optical properties),
where a giant IR resonance is largely separated
from a series of overlapping UV peaks,
by a transparent optional region.
For the UV region, it would be beneficial,
eventually, to experimentally confirm, with updated data sets,
the results reported in
Refs.~\cite{To1936,LeStRo1965} (the latter have been
obtained 88 and 59 years ago, respectively,
and are urgently in need of an update).
Our TDDFT calculations indicate the presence
of multi-excitonic peaks in the energy region
between 10\,eV and 15\,eV [see, in particular,
Fig.~\ref{fig8}(b)]. This result is in agreement with 
the Bethe--Salpeter approach employed in 
Ref.~\cite{BeSh1999} [see the theoretical predictions reported
in Figs.~8, 9 and 10 of Ref.~\cite{BeSh1999}].
An improved understanding of the
UV region, in particular, around the excitonic peak near 10.8\,eV,
would create opportunities for a more detailed
comparison with TDDFT and Bethe--Salpeter 
calculations in the UV. On the experimental side,
the excitonic peak is entirely missing in
Ref.~\cite{BaEtAl1990} but resolved (over a very narrow
spectral range) in Refs.~\cite{ToMi1969,LePaGoRi2003}.

For the time being,
the data published in Ref.~\cite{LeStRo1965}
allow us to devise a consistent model.
We confirm the surprising success of the RRCO model
for the description of the temperature-dependent
dielectric function of technologically important
materials, in the example case of calcium fluoride.

In summary, a careful analysis of the
dielectric function of \CaFT{} reveals
a few, hitherto perhaps not fully appreciated, aspects.
First, one recognizes that the technologically
highly important range from 184\,nm to 2326\,nm
is flanked by pronounced absorptive resonances,
in both the IR region (giant resonance) as well
as the UV region (excitonic resonance).
Second, one realizes that an interpolation between
these regions is possible on the basis of
a RRCO model, which is applicable across the IR, visible,
and UV regions. 

%
%
\section*{Acknowledgments}

The authors acknowledge helpful conversations with
Professor John A.~Leake on general aspects of crystal structure.
T.~D.~and U.~D.~J.~were supported by NSF grants PHY--2110294
and PHY--2513220.
C.~A.~U.~acknowledges support from NSF grant DMR--2149082.

%
%
\section*{Data Availability Statement}

The experimental data used for the dielectric function was extracted from
Refs.~\cite{PSEtAl2009,ZhWaTh2023,LeStRo1965} of the manuscript. Access to the
original publications~\cite{PSEtAl2009,ZhWaTh2023,LeStRo1965}
requires a subscription to the original journal
articles. Theoretical data (fit parameters and results for atom-surface
interactions) are contained in the tables of the article. Other theoretical
data (beyond the figures containing the results of the {\em ab initio} calculations)
that support the findings of this article are not publicly available upon
publication because it is not technically feasible and/or the cost of
preparing, depositing, and hosting the data would be prohibitive within the
terms of this research project. The data are available from the authors upon
reasonable request.

\appendix

\begin{figure}[ht!]
\begin{center}
\includegraphics[width=\figwidthII]{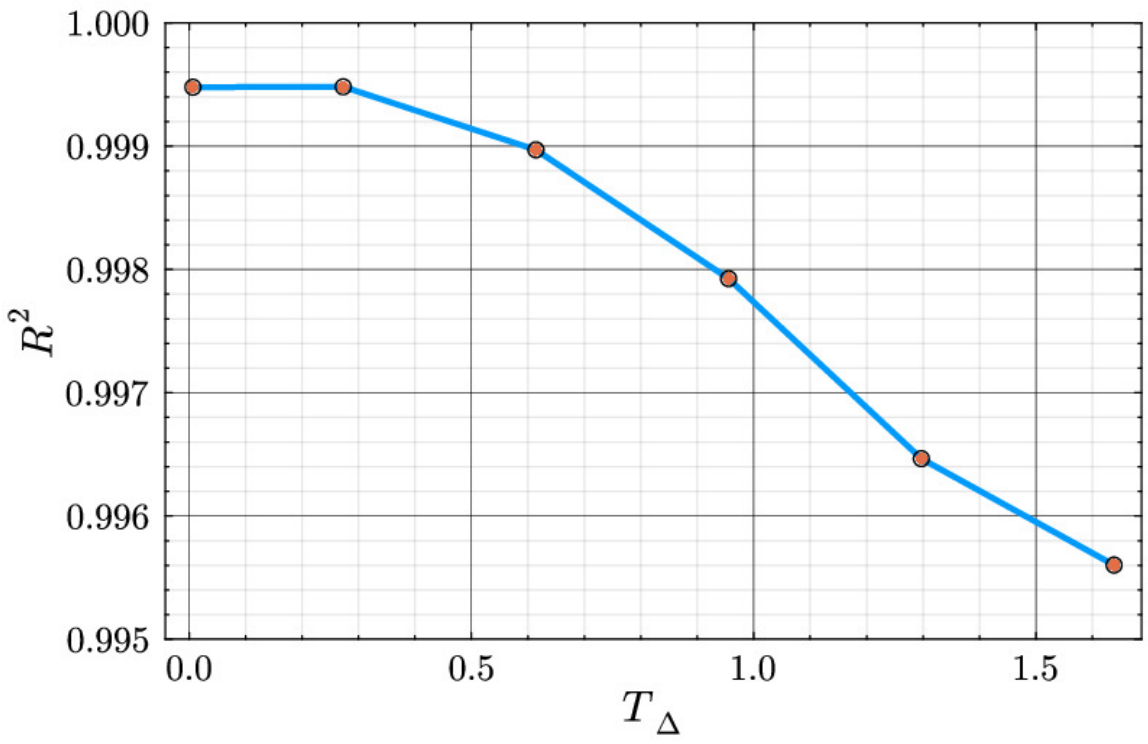}
\caption{
\label{fig14}
Coefficients of determination, $R^2$, plotted as a function of reduced
temperature $T_\Delta$. The results indicate $R^2>0.995$ for all temperatures
under consideration. One notes that the plot range on the
ordinate axis is very condensed and squeezed into the
range $0.995 < R^2 < 1.0$.}
\end{center}
\end{figure}

%
%
\section{Statistical Analysis}
\label{appa}

%
%
\subsection{Parameter Uncertainty}
\label{appa1}

As outlined in Sec.~\ref{sec2C}, our estimate for the accuracy
of our fitting (in the IR and visible range)
is about 5\,\%, which is in line
with the accuracy limits of ellipsometry
measurements for comparable
materials~\cite{MoEtAl2022,MoEtAl2025erratum}.
It is still interesting to compare this estimate
with the results of a standard statistical analysis
of the available data, and of the fits.
We consider the general situation of
data points $(x_i, y_i)$ which need to be fitted to a model
$f(x;\vec{P})$ where $\vec{P}$ is the set of fitting
parameters~\cite{PrFlTeVe1986edition1,PrFlTeVe2007edition3}.
For least-squares-fitting, we define a merit function, $\chi^2$ as
\begin{equation}
\label{eq: chisq}
    \chi^2 = \sum_{i=1}^N \left(\frac{y_i - f(x_i;\vec{P})}{\sigma_i}\right)^2.
\end{equation}
Here, $N$ is the total number of available data points, and $\sigma_i$ denotes
the uncertainty associated with each data point.  Then, we can determine the
best-fit parameters by minimizing $\chi^2$. In order to
estimate the uncertainties in
the fit parameters, we first need to calculate the ``curvature matrix''
$\bm{U}$ which can be defined as follows
[see Eq.~(14.4.8) of
Ref.~\cite{PrFlTeVe1986edition1} and Eq.~(15.5.8) of
Ref.~\cite{PrFlTeVe2007edition3}],
\begin{equation}
    \label{eq: CurvMat}
    U_{mn} = \frac{1}{2} \frac{\partial \chi^2}{\partial P_m \partial P_n}.
\end{equation}
The number of fitting parameters is $M$;
so, the indices $m$ and $n$ assume values $1 \leq m,n \leq M$.
The inverse of $\bm{U}$ is commonly referred to as the
covariance matrix $\bm{C}$. Its off-diagonal elements give the
covariance between the fitting parameters, and the diagonal elements
lead to the standard variances $\sigma^2(P_m)$
of the fitting parameter, $P_m$ [see
Eq.~(14.3.15) of Ref.~\cite{PrFlTeVe1986edition1} and Eq.~(15.4.15)
ff.~of Ref.~\cite{PrFlTeVe2007edition3}].
One has the relation
\begin{equation}
    \sigma^2(P_m) = C_{mm} = \left(U^{-1}\right)_{mm}.
\end{equation}
After carrying out the differentiation with
respect to the parameters in Eq.~\eqref{eq: CurvMat},
we get the following result
[see Eq.~(14.4.7) of Ref.~\cite{PrFlTeVe1986edition1} and
Eq.~(15.5.7) of Ref.~\cite{PrFlTeVe2007edition3}],
\begin{equation}
\begin{split}
U_{mn} &= \sum_{i=1}^{N}  \frac{1}{\sigma_i^2}
\left(
\pdv{f(x_i;\vec{P})}{P_m} \pdv{f(x_i;\vec{P})}{P_n}\right.\\
&\quad\left.
 - (y_i - f(x_i;\vec{P})) \pdv{f(x_i;\vec{P})}{P_m,P_n}
\right).
\end{split}
\end{equation}
The second derivative term is usually ignored in practice because,
for a successful fitting, $(y_i - f(x_i;\vec{P}))$ should be small and randomly
distributed (including its sign). Hence, the second term
should approximately cancel out in the sum. However, the inclusion of
the second derivative term can be destabilizing if there are outliers in the
data. Following Eq.~(14.4.11) of Ref.~\cite{PrFlTeVe1986edition1} and
Eq.~(15.5.11) of Ref.~\cite{PrFlTeVe2007edition3}, we opt to follow the
convention of (re-)defining the matrix $U \rightarrow \bm{U}$ as,
\begin{equation}
\label{eq: CervMat2}
\bm{U}_{mn} = \sum_{i=1}^{N}  \frac{1}{\sigma_i^2}
\left(
\pdv{f(x_i;\vec{P})}{P_m} \pdv{f(x_i;\vec{P})}{P_n}
\right).
\end{equation}
In cases where the uncertainties $\sigma_i$ in the data points are not known, we
can assign a uniform standard deviation $\sigma$ to all data points.
It can be calculated as follows
[see Eq.~(14.1.6) of Ref.~\cite{PrFlTeVe1986edition1} and Eq.~(15.1.7) of
Ref.~\cite{PrFlTeVe2007edition3}],
\begin{equation}
    \sigma^2 = \frac{1}{N_{\rm DOF}}\sum_{i=1}^N \left(y_i - f(x_i;\vec{P})\right)^2,
\end{equation}
where $N_{\rm DOF} = N - M$ is the number of degrees
of freedom. Under these simplifications for
Eq.~\eqref{eq: CervMat2}, we can write
the covariance matrix $\bm{C}$ in terms of
the Jacobian $\bm{J}$ of the model,
\begin{subequations}
\begin{align}
J_{im} =& \; \pdv{f(x_i;\bm{P})}{P_m}, \\
\bm{U}_{mn} =& \; \sum_{i=1}^{N}  \frac{1}{\sigma_i^2}
\left(
\pdv{f(x_i;\vec{P})}{P_m} \pdv{f(x_i;\vec{P})}{P_n}
\right) \nonumber\\
=& \; \frac{1}{\sigma^2} \sum_{i=1}^{N}
\left( \pdv{f(x_i;\vec{P})}{P_m} \pdv{f(x_i;\vec{P})}{P_n}
\right) \,, \\
\bm{U} =& \; \frac{1}{\sigma^2} \bm{J}^T\bm{J} \,, \\
\bm{C} =& \; \sigma^2 \left(\bm{J}^T\bm{J}\right)^{-1} \,.
\end{align}
\end{subequations}
This method of estimating the parameter variance is also mentioned in Eq.~(4.4)
of Ref.~\cite{St2011}.  Finally, we take the square root of the
diagonal elements of $\bm{C}$ to estimate the uncertainties in our fitting
parameters, $\sigma(P_m)$ [no summation over $m$],
\begin{equation}
\sigma(P_m) = \sqrt{C_{mm}} =
  \sigma \sqrt{\left(\left(\bm{J}^T\bm{J}\right)^{-1}\right)_{mm}}.
\end{equation}
The results for the
uncertainties in the fitting coefficients are presented
(for $T = 22 \, \C$) in Table~\ref{table4}
and (for $T = 500\, \C$) in Table~\ref{table5}.
The entries for the IR peak ($k=1$) have a much smaller
uncertainty than those for the UV peaks.
This is natural because the IR peak
has been measured to excellent
accuracy in Ref.~\cite{PSEtAl2009},
while our knowledge of the UV region is
not as detailed~\cite{LeStRo1965}.

%
%
\subsection{Coefficient of Determination}
\label{appa2}

One might ask if our fit has been successful.
A general measure of the quality of the fit is the
coefficient of determination, commonly referred
to as $R^2$.
The most general definition of $R^2$ is given as
Table 1.5, page 15 of Ref.~\cite{RaPaDi2001}
\begin{subequations}
\begin{align}
R^2 &= 1 - \frac{SS_{\rm res}}{SS_{\rm tot}},\\
SS_{\rm res} &= \sum_{i}(y_i - f(x_i))^2,\\
SS_{\rm tot} &= \sum_{i}(y_i - \bar{y})^2 \,.
\end{align}
\end{subequations}
Here, $SS_{\rm res}$ is the ``residual sum of squares'',
and $SS_{\rm tot}$ is the ``total sum of squares''.
Furthermore, $\bar{y}$ is the mean of $\{y_i\}$.
For all temperatures considered in our fitting,
we obtain values of $R^2 > 0.995$ as presented in Fig.~\ref{fig14}.
A coefficient of determination approximately
equal to unity implies that the fitting procedure
was successful.

\vspace*{0.3cm}

%
%
\section{Crystal Structure, Symmetries and Limiting Factors}
\label{appb}

Single crystals of CaF$_2$ have a face-centered cubic
(fcc) structure, 
where standard cleavage occurs along the (111) planes.
We now take the opportunity to explicitly point out
some approximations used in the current study,
for absolute clarity.
Firstly,
the space group, ${\rm Fm} {\bar 3} {\rm m}$,
and the point group
$O_h({\rm m} {\bar 3} {\rm m})$, of calcium fluoride are
well known (see Ref.~\cite{Ny1985}).
Our investigations use the
dipole approximation for the dielectric tensor
(see Refs.~\cite{PaVe1971,DeDaChGo1985}).
Within this approximation,
the cubic point group of the crystal implies
an identical refractive index in all crystal planes,
notably, the (111) and (100) planes.
The dipole approximation becomes exact in the
long-wavelength limit (see also Chap.~6 of
Ref.~\cite{Je2017book} and Ref.~\cite{Ru1970}).
We thus do {\em not} consider the numerically
small intrinsic birefringence (of order $10^{-7}$) which can be
observed in the UV and theoretically explained
by the breaking of cubic symmetry for interactions
at short wavelengths in the ultraviolet
(see Refs.~\cite{PaVe1971,DeDaChGo1985,BuLeSh2001,BuLeShBr2001}).
We also do {\em not} consider
conceivable strain-induced birefringence
effects~\cite{RaLeWi1995,dataCorningHellma,CaF2corning,CaF2hellma},
nor potential dislocation-induced modifications
of the dielectric function~\cite{Co1961,ZhEtAl2025}.

A further aspect beyond the approximations used
in the current paper would concern
conceivable, crystal-plane-specific surface
rearrangements of the crystal structure
in the surface layers; such rearrangements
have been observed for the (111) plane of intrinsic
silicon~\cite{BiRoGeWe1982}, affecting the two
uppermost surface layers.
For wavelengths large against the dimensions of the
unit cell of the crystal, such rearrangements are
expected to have numerically very small effects
on the dielectric function.

Finally, we emphasize that, of course, the
treatment of the temperature-dependence
of the atom-surface interaction is restricted to
the explicit temperature dependence
of the dielectric function of the material.
A further temperature-dependent effect,
due to the ``thermal discretization'' of the frequencies
of the virtual photons that mediate the atom-surface
interaction, in terms of the Matsubara
frequencies~\cite{DzLiPi1959jetp,DzLiPi1961spu,DzLiPi1961advphys},
is {\em not} considered here.

We also clarify that we restrict the discussion to the atom-surface
interaction of ground-state atoms, and those in very long-lived metastable
states (metastable helium). This has some implications. Namely, for
interactions of excited (not ground-state) Cs atoms with CaF$_2$ and BaF$_2$
surfaces, due to the close spectral proximity of certain dipole-allowed atomic
transitions and surface polaritons, a resonant enhancement of the temperature
dependence of the atom-surface interaction can be expected~\cite{PSEtAl2014}. 
We have checked that there is no such spectral overlap for metastable helium in
the $2 {}^3 S_1$ state. Namely, for metastable helium in the triplet state, the
lowest dipole-allowed transitions to excited $2 {}^3 P_{J=0,1,2}$ states are,
energetically, at least 20 times higher than the surface 
polariton~\cite{PSEtAl2014,NISTASD_2024}.

\begin{figure}[t!]
\begin{center}
\includegraphics[width=0.91\linewidth]{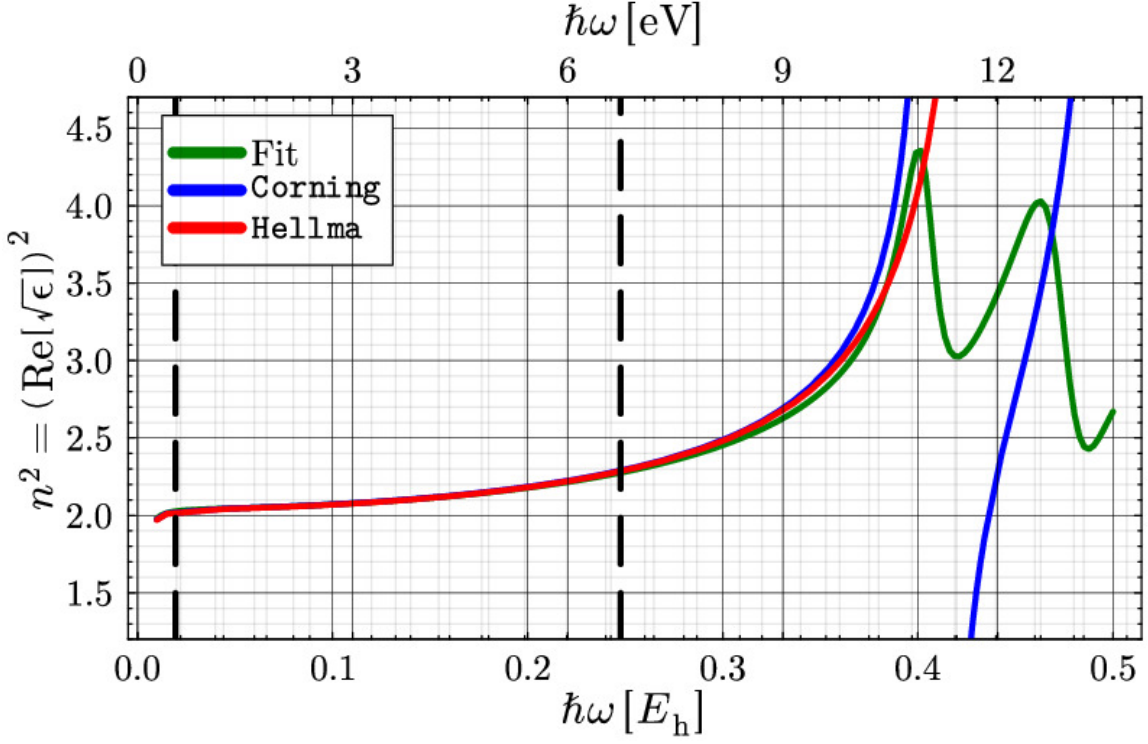}
\caption{ \label{fig15}
The square of the refractive index, $n^2 = [\Re(\sqrt{\epsilon})]^2$,
is compared between the
{\sc Hellma} data sheet (Ref.~\cite{CaF2hellma}, red curve), the
{\sc Corning} data sheet (Ref.~\cite{CaF2corning}, blue curve),
and the calculations reported here (green curve).
The vertical bars are located at $\omega = \omega_{\rm L} $
and $\omega = \omega_{\rm H}$ [see Eq.~\eqref{omegaLH}].}
\end{center}
\end{figure}

%
%
\section{Comparison to Commercial Data Sheets}
\label{appc}

Commercially available
data sheets~\cite{CaF2corning,CaF2hellma}
contain formulas for the refractive index
valid in the wavelength range
$184 \, {\rm nm} < \lambda < 2326 \, {\rm nm}$,
which translates into the frequency interval
\begin{equation}
\label{omegaLH}
\omega_{\rm L} \equiv
0.019588 \, {\rm a.u.} < \omega <
\omega_{\rm H} \equiv
0.247627 \, {\rm a.u.} \,.
\end{equation}
The {\sc Hellma} data sheets~\cite{CaF2hellma}
contain the formula
\begin{align}
\label{nH2}
n_{\rm H}^2 =& \; 1 + \sum_{i=1}^3 \frac{a_{{\rm H},i}}{\omega_{{\rm H},i}^2 - \omega^2} \,.
\end{align}
The parameters correspond to three undamped oscillator
peaks with the resonance frequencies outside of the
interval $\omega \in (\omega_{\rm L}, \omega_{\rm H})$,
\begin{align}
a_{{\rm H},1} =& \;  6.6594 \times 10^{-6} \,, &  \omega_{{\rm H},1} & = 0.0013941 \,, \\
a_{{\rm H},2} =& \;  0.08214 \,,  & \omega_{{\rm H},2} & = 0.44229 \,, \\
a_{{\rm H},3} =& \;  0.46548 \,,  & \omega_{{\rm H},3} & = 0.86730 \,.
\end{align}
outside of the interval $\omega \in (\omega_{\rm L}, \omega_{\rm H})$,
The {\sc Corning} data sheets~\cite{CaF2corning}
(for $T = 20^\circ \, {\rm C}$, or $T = 293 \, {\rm K}$)
correspond to four undamped oscillator peaks
outside of the interval $\omega \in (\omega_{\rm L}, \omega_{\rm H})$,
\begin{align}
\label{nC2}
n_{\rm C}^2 =& \; 1 + \sum_{i=1}^4 \frac{a_{{\rm C},i}}{\omega_{{\rm C},i}^2 - \omega^2} \,.
\end{align}
The entries for the four resonance peaks in the 
{\sc Corning} data~\cite{CaF2corning} are
\begin{align}
a_{{\rm C},1} =& \;  6.6812  \times 10^{-6} \,, &  
\omega_{{\rm C},1} & = 0.00086868 \,, \\
a_{{\rm C},2} =& \;  0.025547 \,, & \omega_{{\rm C},2} & = 0.41062 \,, \\
a_{{\rm C},3} =& \;  0.11481 \,,  & \omega_{{\rm C},3} & = 0.51035 \,, \\
a_{{\rm C},4} =& \;  0.52877 \,,  & \omega_{{\rm C},4} & = 1.08850  \,.
\end{align}
In order to compare the results to our calculations,
we observe that, if $n + \ii \kappa  = \sqrt{ \epsilon }$, then
\begin{align}
n =& \; \Re \, \sqrt{\epsilon} = \frac{1}{\sqrt{2}} \,
\left[ \sqrt{ [\Re(\epsilon)]^2 + [\Im(\epsilon)]^2} + \Re(\epsilon) \right]^{1/2} \,, \\
\kappa =& \; \Im \, \sqrt{\epsilon} = \frac{1}{\sqrt{2}} \,
\left[ \sqrt{ [\Re(\epsilon)]^2 + [\Im(\epsilon)]^2} - \Re(\epsilon) \right]^{1/2} \,.
\end{align}
The comparison in Fig.~\ref{fig15} reveals that excellent agreement
between our calculations and data sheets~\cite{CaF2corning,CaF2hellma}
is achieved in the interval $\omega \in (\omega_{\rm L}, \omega_{\rm H})$,
while, outside of this interval, the formulas from the
data sheets~\cite{CaF2corning,CaF2hellma}
are clearly not applicable, in view of divergences caused by the
undamped oscillator peaks.

%
%
\section{RRCO Model for Rutile}
\label{appd}

\begin{table}[t!]
\centering
\caption{ \label{table8}
Parameters for rutile (TiO$_2$) taken from Table II of Gervais and Piriou
(Ref.~\cite{GePi1974})
for the ordinary axis, for room temperature.
Units are ${\rm cm}^{-1}$. For the conversion to 
atomic units, we have used the fact that one inverse
centimeter corresponds to an energy of 
$1.23985 \times 10^{-4} \, {\rm eV}$ (see Appendix~14 
of Ref.~\cite{BrJo2000}).}
\setlength{\tabcolsep}{3.5pt}
\begin{tabular}{lcccc}
\toprule
\toprule
$j$ &
\multicolumn{1}{c}{$\Omega_{j \TO}~[{\rm cm}^{-1}]$} &
\multicolumn{1}{c}{$\gamma_{j \TO}~[{\rm cm}^{-1}]$} &
\multicolumn{1}{c}{$\Omega_{j \LO}~[{\rm cm}^{-1}]$} &
\multicolumn{1}{c}{$\gamma_{j \LO}~[{\rm cm}^{-1}]$} \\
\midrule
1  & $189.0$ & $27.0$ & $831.0$ & $50.0$ \\
2  & $381.5$ & $16.5$ & $367.0$ & $10.0$ \\
3  & $508.0$ & $24.0$ & $443.5$ & $21.5$ \\
4  & $585.5$ & $65.0$ & $575.0$ & $65.0$ \\
\bottomrule
\bottomrule
\end{tabular}
\end{table}

In order to illustrate the wide applicability of the 
RRCO model, we briefly review its relation 
to the so-called four-parameter semiquantum (FPSQ) 
model~\cite{GePi1974,AmEtAl2023},
which had been used in Ref.~\cite{GePi1974} 
in order to describe the dielectric function
of rutile (TiO\textsubscript{2}) along the ordinary 
and extraordinary axes. We compare our {\em ansatz} 
to the FPSQ model by way of example.
Parameters used in Ref.~\cite{GePi1974} for the 
ordinary axis, at room temperature, are given
in Table~\ref{table8}. The four-term,
factorized form of the dielectric function of the FPSQ model
is given as
\begin{equation}
\label{eq:rutile}
\epsilon_\mathrm{FPSQ}(\omega)
= \epsilon_\infty \prod_{j=1}^4
\frac{\Omega_{j\mathrm{LO}}^2 - \omega^2 - \ii \gamma_{j\mathrm{LO}}\,\omega}
{\Omega_{j\mathrm{TO}}^2 - \omega^2 - \ii \gamma_{j\mathrm{TO}}\,\omega}\,,
\end{equation}
where a value of $\epsilon_\infty = 6.0$ was used by 
the authors of Ref.~\cite{GePi1974} for the ordinary axis.
The FPSQ model given in Eq.~\eqref{eq:rutile} is inspired by
Lyddane–-Sachs–-Teller (LST) relation~\cite{AsMe1976},  and is thus
applicable in the far-IR and mid-IR region~\cite{ScEtAl2013} to
characterize the phonon modes~\cite{LaLiJu2018}.  In LST-type models,
contributions from higher frequencies due to interband transitions are
approximated by choosing an appropriate value of the $\epsilon_\infty$,
which can be different from unity.
(Thus, $\epsilon_\infty$ is not equal to the 
true limit of the dielectric function for high frequencies,
which reads as
$\lim_{\omega \to \infty} \epsilon(\omega) = 1$.)
In particular, the FPSQ model with $\epsilon_\infty = 6.0$
cannot fulfill the Kramers--Kronig relations because 
the integral which describes its imaginary part 
as a  function of its real part does not 
converge in the high-energy limit [see Eq.~(6.207b) 
of Ref.~\cite{Je2017book}].
The far-IR and mid-IR region thus represents a ``cutoff''
for the validity of the FPSQ model.
In order to obtain a functional form
consistent with the Kramers--Kronig relations, the data from 
Ref.~\cite{GePi1974} would need to be supplemented by data in the UV region 
({\em e.g.}, from sources such as Ref.~\cite{JeCuAr2024}) as the band gap of
rutile is at the upper end of the visible spectrum
($3\,{\rm eV} \approx 0.1 \,{\rm a.u.}$).
The completion of this endeavor is left for future investigations.

It is extremely interesting to observe that, by a partial-fraction 
decomposition, we can write the FPSQ model in an exactly 
equivalent form which, up to a generalization of the 
constant term, $1 \to \epsilon_\infty$, matches the 
functional form of the RRCO model, 
\begin{equation}
\label{master2}
\epsilon_\mathrm{FPSQ}(\omega)
= \epsilon_\infty +  \sum_{k=1}^{4}
\frac{a_k(\omega_k^2 - \ii \gamma'_k \omega)}%
{\omega_k^2 - \omega^2 - \ii \gamma_k\omega},
\end{equation}
where the relevant parameters are given in Table~\ref{table9}.

\begin{table}[t!]
\centering
\caption{\label{table9}
Parameters entering the FPSQ model outlined in 
Eq.~\eqref{eq:rutile}, after conversion to the 
RRCO form~\eqref{master2}.}
\begin{ruledtabular}
\begin{tabular}{crccr}
{$k$} & 
\multicolumn{1}{c}{$a_k$} & 
\multicolumn{1}{c}{$\omega_k~[E_h/\hbar]$} & 
\multicolumn{1}{c}{$\gamma_k~[E_h/\hbar]$} & 
\multicolumn{1}{c}{$\gamma'_k~[E_h/\hbar]$} \\
\midrule
1 & 69.0717 & $0.8611 \times 10^{-3}$ & $1.230 \times 10^{-4}$ & $-9.198 \times 10^{-6}$ \\
2 &  0.9307 & $1.7382 \times 10^{-3}$ & $7.518 \times 10^{-5}$ & $5.404 \times 10^{-4}$ \\
3 &  2.6360 & $2.3146 \times 10^{-3}$ & $1.094 \times 10^{-4}$ & $2.766 \times 10^{-4}$ \\
4 &  0.4034 & $2.6655 \times 10^{-3}$ & $2.962 \times 10^{-4}$ & $-5.301 \times 10^{-4}$ \\
\end{tabular}
\end{ruledtabular}
\end{table}

\color{black}


\begin{thebibliography}{95}
\bibitem [{dat()}]{dataCorningHellma}%
  \BibitemOpen
  \href@noop {} {}\bibinfo {howpublished} {Strain-optic coefficients for
  CaF$_2$ can be found on pertinent web pages of Corning and Hellma companies
  (see Refs.~\cite{CaF2corning} and~\cite{CaF2hellma}).}\BibitemShut {Stop}%
\end{thebibliography}
\end{document}